\documentclass[sigconf,anonymous=false]{acmart}

\renewcommand\footnotetextcopyrightpermission[1]{} 
\usepackage{graphicx}
\usepackage{balance}  
\usepackage{xspace}
\usepackage{xcolor}
\usepackage{alltt}
\usepackage{listings}
\usepackage{color}
\usepackage{caption}
\usepackage{subcaption}
\usepackage{tabularx, tabulary}
\usepackage{makecell}
\usepackage{booktabs}
\usepackage[noend]{algpseudocode}
\usepackage{algorithm}
\usepackage{multirow}
\usepackage{pifont}
\usepackage{enumitem}
\usepackage{mdframed,lipsum}

\newcommand{\TODO}[1]{\textcolor{orange}{TODO: #1}}

\newcommand{\rev}[1]{\textcolor{black}{#1}}
\newcommand{\pb}[1]{\textcolor{black}{#1}}

\newcommand{\heading}[1]{\vspace{4pt}\noindent\textbf{#1}}

\DeclareRobustCommand{\TODO}[1]{\textcolor{orange}{TODO: #1}}

\makeatletter
\def\@copyrightspace{\relax}
\makeatother

\definecolor{codegreen}{rgb}{0,0.6,0}
\definecolor{codegray}{rgb}{0.5,0.5,0.5}
\definecolor{codelightgray}{rgb}{0.5,0.5,0.5}
\definecolor{codepurple}{rgb}{0.58,0,0.82}
\definecolor{backcolour}{HTML}{FFFFE6}
\definecolor{uoftblue}{rgb}{0, 0.16, 0.36}
\definecolor{rulecolor}{HTML}{6D5000}
\definecolor{darkred}{HTML}{9B2036}

\lstdefinestyle{qzstyle}{
    backgroundcolor=\color{backcolour},
    commentstyle=\color{codelightgray},
    keywordstyle=\color{darkred},
    numberstyle=\tiny\color{codegray},
    stringstyle=\color{darkred},
    basicstyle=\ttfamily\footnotesize,
    breakatwhitespace=false,
    breaklines=true,
    captionpos=b,
    keepspaces=false,
    numbers=left,
    numbersep=5pt,
    showspaces=false,
    showstringspaces=false,
    showtabs=false,
    tabsize=2,
    frame=ltb,
    frame=single
}

\mdfdefinestyle{FindingFrame}{%
    linecolor=black,
    outerlinewidth=2pt,
    innertopmargin=4pt,
    innerbottommargin=4pt,
    innerrightmargin=4pt,
    innerleftmargin=4pt,
        leftmargin = 4pt,
        rightmargin = 4pt,
    backgroundcolor=gray!20!white
}

\begin{document}

\setlength{\abovedisplayskip}{3pt}
\setlength{\belowdisplayskip}{3pt}

\newcommand{\name}[0]{{\textsc{dpBento}}\xspace}
\newcommand{\octeon}[0]{{\textsc{OCTEON}}\xspace}
\newcommand{\nvbf}[0]{{\textsc{BF-2}}\xspace}
\newcommand{\nvbff}[0]{{\textsc{BF-3}}\xspace}

\title{\name: Benchmarking DPUs for Data Processing}

\author{Jiasheng Hu, Chihan Cui, Anna Li, Raahil Vora, Yuanfan Chen,\\\vspace{-3mm} Philip A. Bernstein$^\dag$, Jialin Li$^\S$, Qizhen Zhang\\University of Toronto, $^\dag$Microsoft Research, $^\S$National University of Singapore}

\begin{abstract}

Data processing units (DPUs, SoC-based SmartNICs) are emerging data center hardware that provide opportunities to address cloud data processing challenges.
Their onboard compute, memory, network, and auxiliary storage can be leveraged to offload a variety of data processing tasks.
Although recent work shows promising benefits of DPU offloading for specific operations, a comprehensive view of the implications of DPUs for data processing is missing.
Benchmarking can help, but existing benchmark tools lack the focus on data processing and are limited to specific DPUs.
In this paper, we present \name, a benchmark suite that aims to uncover the performance characteristics of different DPU resources and different DPUs, and the performance implications of offloading a wide range of data processing operations and systems to DPUs.
It provides an abstraction for automated performance testing and reporting and is easily extensible.
We use \name to measure recent 
DPUs, present our benchmarking results, and highlight insights into the potential benefits of DPU offloading for data processing.

\end{abstract}

\maketitle

\section{Introduction}
\label{sec:intro}

Data processing workloads, e.g., databases~\cite{aurora, socrates, alloydb, polardb} and KV stores~\cite{rocksdb_fb, dynamodb, dynamodb2, azure_redis}, 
hosted in cloud data centers are growing.
So are the challenges.
First, while computing demand for data processing keeps increasing due to data growth, general-purpose processors (CPUs) cannot sustain high compute efficiency due to the slowdown of hardware scaling laws.
Hardware resource disaggregation (e.g., that of storage~\cite{Snowflake20, Taurus20, rack_store_disagg} and memory~\cite{QizhenCIDR20, QizhenVLDB20, disaggdbms_tut, farview}) also decouples the software components of cloud-native systems, leading to intensive network communication activities, which have become the primary performance bottleneck~\cite{aurora}.
In addition, the speeds of storage and network devices are rapidly advancing (e.g., NVIDIA ConnectX-7 NICs provide 400\,Gbps bandwidth~\cite{cx7}).
Since CPU instructions consumed to perform one byte of I/O are fixed, increases in disk and network bandwidth lead to high CPU expenses~\cite{dana_cidr20, cowbird}.

Data processing units (DPUs) are SoC-based SmartNICs that have emerged as popular programmable devices for in-network offloading.
A DPU is characterized with a set of hardware resources curated to optimize data-path efficiency, including (1) CPU cores based on energy-efficient architectures, e.g., Arm and MIPS, (2) memory of moderate size, typically from 8\,GB to 32\,GB, (3) ASIC accelerators for specific compute-intensive tasks such as compression and encryption, (4) high-speed network interface that provides 100s of Gbps bandwidth, and (5) PCIe access to host resources and peer devices.
These resources are promising solutions to the above challenges \rev{in cloud data processing}.
Specifically, CPU-intensive tasks (for both computational operators and I/O) can be offloaded to hardware accelerators and DPU cores to improve compute efficiency and reduce host CPU utilization; 
high-speed network connectivity and optimized I/O libraries can be leveraged to achieve better data movement performance over the network.

Current DPUs are designed for offloading low-level packet processing, storage protocols, and network security.
\rev{{\em Are DPUs also useful for cloud data processing?} 
Recent work has shown concrete benefits of DPU offloading for 
databases and KV stores~\cite{dds, smartshuffle, ThostrupFZB22}.
However, these proposals have only explored limited DPU capabilities, such as userspace networking, onboard DRAM, and power-efficient CPUs.
They have been applied to only a few operation types, such as
remote storage reads~\cite{dds}, indexing~\cite{ThostrupFZB22}, and shuffling~\cite{smartshuffle}.
Moreover, they only targeted one DPU product.} 

Another line of work benchmarks DPUs from {\em specific vendors} for {\em specific tasks}.
For instance, DPU-bench~\cite{dpu-bench} measures the efficiency of MPI communication between DPUs, DPUBench~\cite{dpubench} provides a benchmark suite for NVIDIA BlueField-2 DPU (BF-2), and Wei et al.~\cite{bf2_rdma} and Liu et al.~\cite{bf2_perf_char} evaluated RDMA and computational efficiency of BF-2, respectively.
None of these studies target \rev{data processing workloads}.
Moreover, each benchmark targets one device type, overlooking the large hardware space, such as
NVIDIA BlueField~\cite{nvidia_bluefield}, AMD Pensando~\cite{amd_pensando}, Intel IPU~\cite{intel_ipu}, Marvell OCTEON~\cite{marvell_octeon}, and AWS Nitro~\cite{aws_nitro}. 

\rev{We believe a broader DPU benchmark for data processing workloads is needed.
This can help us understand how existing findings 
can be generalized to DPUs from different vendors and across different generations.
The benchmark should include different data processing tasks, exercise different hardware resources on DPUs, and be portable to different DPU devices.
Benchmark results can provide insights into which data processing workloads are suitable for DPU offloading, and whether to offload
primitive instructions, database components, or an end-to-end system.}

To fill this need, we present \name, a benchmark \rev{framework} that automates \rev{the generation, execution, and reporting of data processing performance measurement tests} on DPUs
We use \name to present benchmark results on multiple mass-production DPUs.

\heading{Benchmark framework.}
\name provides a {\em task abstraction} for specifying the workflow of \rev{extensive} measurement tests on a DPU: preparing the environment, executing tests according to user configurations, generating an informative report given metrics of interest, and finally cleaning up the DPU by removing any effects of the measurement.
Based on this \rev{general} abstraction, we developed \rev{built-in tests 
for (1) primitive operations} closely related to data processing efficiency, including tasks that measure the performance of compute, memory, storage, and networking\rev{, (2) cloud database system modules that verify the benefits of DPUs, including predicate pushdown and index offloading, and (3) macro database workloads with a popular lightweight DBMS (DuckDB)}.
A task can be configured to generate different tests.
For example, the compute task is parameterized to test 
\rev{arithmetic or string operations 
---computational operations commonly executed in data systems.}

To use \name, a user simply creates a configuration file, which we call a {\em measurement box} or {\em box} for short.
For each measurement, the box defines tasks, their parameters,
and the performance metrics of interest.
For instance, a user can declare a box for measuring the latency of 
reading memory objects 
and 
\rev{the throughput of table scans with predicate pushdown.}
Given a box configuration, \name  automatically generates tests, executes the workflow of each task, collects measurement results, and eventually presents the test results to the user.
\rev{This unifies} DPU performance tests for various data processing tasks into a single framework. 

A major goal of \name 
is extensibility.
DPUs use product-specific accelerators.
For instance, NVIDIA BF-2 is equipped with a compression ASIC that is missing in BF-3 or Marvell \octeon 10.
The SDKs for programming accelerators are also vendor-specific, such as Data-center-On-a-Chip Architecture (DOCA) for BlueField~\cite{doca} and extensions to base Linux SDK for \octeon~\cite{octeon_sdk}.
To measure the effects of hardware acceleration and support ad-hoc data processing tasks (e.g., 
\rev{specific data system modules} a user considers offloading to the DPU),
\name allows {\em plugin tasks}.
These tasks are implemented by users following the task abstraction and dropped into the framework.
When performing a measurement, the user can declare tests in a plugin task 
in the box configuration.
These tests will then be generated and executed by \name.
As a benefit of the modular design of \name, plugins enable more extensive data processing benchmarking on DPUs. 


\heading{Benchmarking mass-production DPUs.}
Utilizing \name, we perform a comprehensive evaluation of popular DPUs for data processing workloads.
\name's built-in 
\rev{tasks cover a range of workloads from primitive operations, to database system modules, and to a full DBMS}.
\rev{We run them all} to measure DPUs from NVIDIA (BlueField-2 and BlueField-3) and Marvell (OCTEON TX2).
We also implement 
plugin tasks to assess the improvement of hardware accelerators on these DPUs for compute-intensive operations and the benefits of userspace networking.

We analyze and present the experimental results.
If a test can measure multiple DPUs, e.g., a built-in 
\rev{primitive operation} test, we compare its results across these DPUs and the host.
With plugin tasks that work only on specific DPUs, we evaluate the benefits of DPU-specific features. 
Our 
analysis reveals 
insights into DPU offloading for data processing. 
For instance, while the CPU cores on DPUs are weaker for general applications, they outperform host CPUs for certain operations; DPU hardware accelerators can achieve exceedingly high throughput but \rev{at the cost of high latency}; \rev{and rather than run a full DBMS on a DPU, offloading database sub-modules can better exploit DPU capabilities.} 

In summary, this paper makes the following contributions:
\begin{itemize}
    \item Proposes \name, a unified benchmark suite for measuring data processing tasks centered on DPUs ($\S$\ref{sec:overview}). 
    \item Benchmarks recent mass-production DPUs ($\S$\ref{sec:hardware}). 
    \item Presents the benchmark results and provides insights into the compute and I/O efficiency ($\S$\ref{sec:compute-efficiency}---$\S$\ref{sec:io-efficiency})\rev{, the benefits of in-network offloading of cloud database modules ($\S$\ref{sec:module}), and the overhead of running a full DBMS ($\S$\ref{sec:e2e}) on the DPUs.} 
\end{itemize}

\section{Background}
\label{sec:background}

\subsection{\rev{Data Processing Units}}
\label{sec:dpu}


DPUs are programmable network interface cards (NICs) or SmartNICs that allow for data-path customizations and optimizations based on System-on-a-Chip (SoC) architectures.
The target data-path operations include packet processing and disaggregated storage protocols in data center networks.

\begin{figure}[t]
    \centering
        \centering
        \includegraphics[width=\columnwidth]{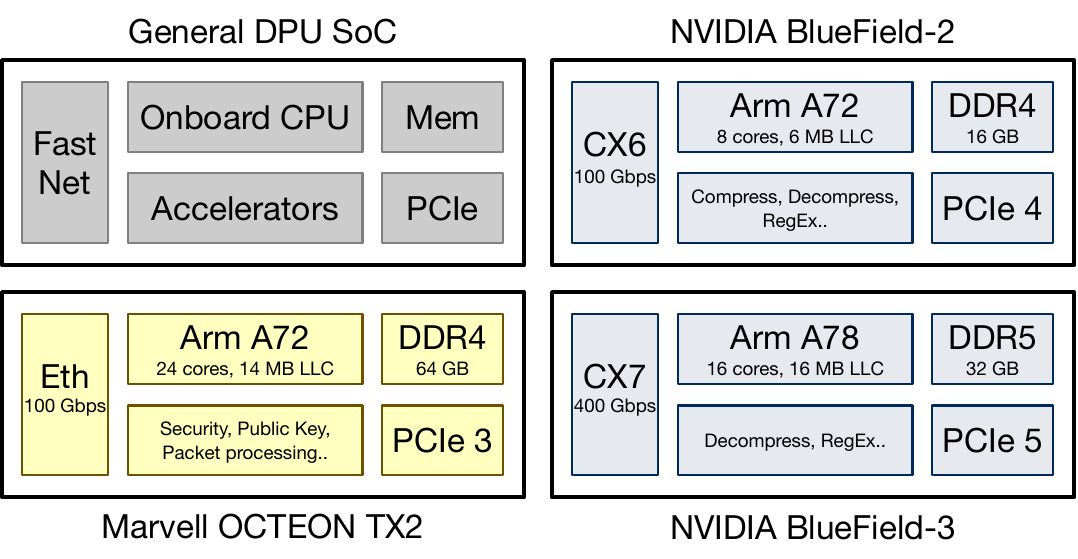}
        \caption{DPU architecture and DPUs from different vendors.}
        \label{fig:dpu-arch}
        \vspace{-3mm}
\end{figure}

\pb{DPUs can be characterized by the hardware resources they support (see Figure~\ref{fig:dpu-arch}).}
As a NIC, a DPU usually provides state-of-the-art network speed, achieving 100s of Gbps bandwidth, 
for fast data movement between DPUs.
To provide ease of programming and access, a DPU is equipped with an onboard CPU and a standard OS, e.g., Linux.
For energy efficiency the CPU adopts a low-power architecture, e.g., Arm or MIPS.
To accelerate popular data-path tasks that are compute-intensive, e.g., compression, encryption, and regular expression (RegEx) matching, a DPU hardens these tasks as onboard accelerators to achieve ultra high performance and energy efficiency.
A DPU also has a moderate amount of DDR memory to hold the working set of offloaded applications, typically ranging from 8\,GB to 32\,GB.
A PCIe switch on a DPU gives it access to external system-wide resources, e.g., host memory and peer PCIe devices such as SSDs and GPUs.
While not essential for low-latency high-throughput data-path applications, a DPU needs auxiliary local storage, e.g., 
an SSD, to install firmwares and the OS. 

Compared to ASIC-based and FPGA-based SmartNICs, DPUs offer general programmability with onboard CPUs, and higher compute performance and energy efficiency with hardware accelerators.
Given current trends in cloud infrastructure such as virtualization~\cite{accelnet}, high-performance I/O~\cite{dana_cidr20, nvmeof}, and disaggregation~\cite{dds, net_req_disagg}, DPUs are becoming a popular solution to offload host functionalities to lower \rev{TCO} and improve data-path efficiency~\cite{dpu_deploy_alibaba, dpu_deploy_azure_amd, dpu_deploy_vmware, dpu_deploy_nitro}.
Recent work has shown promising results for offloading certain data processing tasks to DPUs~\cite{dds, dpdpu, ThostrupFZB22}.

\subsection{DPU Variety}
A crucial consideration of DPU offloading for data processing is the variety of DPUs, which is reflected in the following aspects.
\begin{itemize}[leftmargin=3mm]
    \item {\em Various resources.} 
    \pb{A DPU SoC has resources for general and specialized computation},
    memory, storage, and interconnections to the network and to other resources on the host.
    All these resources can be utilized by a data processing system but present different performance characteristics compared to host servers.
    \item {\em Various vendors.}
    \pb{Many commercial DPUs are being produced}:
    NVIDIA BlueField~\cite{nvidia_bluefield}, Marvell OCTEON~\cite{marvell_octeon},
    Intel IPU~\cite{intel_ipu}, AMD Pensando~\cite{amd_pensando}, Broadcom Stingray~\cite{broadcom_stingray}, Kalray MPPA~\cite{kalray_mppa}, AWS Nitro~\cite{aws_nitro}, Alibaba CIPU~\cite{alibaba_cipu}, and Microsoft Fungible~\cite{azure_fungible}. 
    \pb{And most cloud providers are deploying them.}
    Figure~\ref{fig:dpu-arch} shows the specs of NVIDIA  BlueField-3 and Marvell OCTEON TX2, two state-of-the-art DPUs we benchmark in this work.
    Although they follow the same high-level architecture, detailed hardware configurations, e.g., the network interface and 
    \pb{hardware accelerators}, greatly differ \pb{(details are in $\S$\ref{sec:hardware})}.
    Moreover, DPU SDKs are often vendor-specific.
    For instance, NVIDIA provides the Datacenter-On-a-Chip Architecture (DOCA) SDK~\cite{doca} to program BlueField DPUs, while Marvell offers more standard Linux toolchains and Data Plane Development Kit (DPDK) for OCTEON~\cite{octeon_sdk}.
    \item {\em Various generations.} Even for the same vendor, DPUs \rev{across} generations can be configured differently.
    For instance, from BF-2 to BF-3, not only are the CPU, memory, PCIe, and NIC upgraded, but the set of hardware accelerators has also been changed.
\end{itemize}

\pb{Creating and running a general benchmark suite 
for} state-of-the-art DPUs are keys to understanding the implications of different DPU resources and different DPUs on cloud data processing.
\section{{\name} Framework}
\label{sec:overview}

We present \name, a DPU-centric benchmark framework.
It has four primary goals: (1) {\em extensibility} such that it can implement tests to stress different DPU resources for data processing, (2) {\em customizability} such that it allows users to focus on specific data processing aspects with selective tests, (3) {\em portability} such that it can measure different DPUs, and (4) {\em readiness} such that its off-the-shelf repository already contains a range of tests that users can run to benchmark their DPUs.
This section details the extensible abstraction, customizable measurement, portable workflow, and microbenchmark and system \rev{module} tests that we provide to evaluate a wide spectrum of data processing operations and workloads.

\subsection{Task Abstraction}
In \name, a 
data processing workload is implemented as a {\em task}.
\pb{It can be a primitive operation such as memory reads or a coarse-grained database module such as predicate pushdown. 
A task is parameterized by {\em performance tests} to evaluate a DPU SoC with {\em metrics of interest}
and is executed following the four steps below.}

\begin{itemize}
    \item {\bf Prepare.} Executing performance tests on a DPU may involve installing external libraries and toolkit as dependencies, downloading open datasets, and setting up directories.
    This step is to properly prepare the DPU environment. 
    \item {\bf Run.} With concrete parameters that specify a test (e.g., random reads to memory) 
    and target performance metrics (e.g., average latency), this step is to carry out the performance test and generate logs that contain the measurement result.
    \item {\bf Report.} The results of individual tests can be temporarily cached in the logs.
    This step is to process the intermediate logs to produce a report that is organized, formatted, and finally presented to the user.
    \item {\bf Clean.} Benchmarking a DPU should not modify its state---no permanent effect is expected or allowed. This step is to clean all the directories, files, and software and system configurations generated in this task to restore the DPU to its state before running this task.
\end{itemize}

These steps constitute the interface of \name.
To implement a task, 
\pb{a user} provides the logic for each step as a Python script.
The framework will configure and execute it \pb{as specified.}
Based on this abstraction, we have implemented a wide range of data processing performance tests as tasks, including microbenchmarks ($\S$\ref{design:micro}), \rev{cloud database modules ($\S$\ref{design:module}), and a full DBMS ($\S$\ref{design:macro})}.


\subsection{\rev{Extensibility}}
Users can
\rev{customize their DPU performance testing as well as extend the benchmark suite.}
First, a performance measurement job can be composed with a customized set of tests from 
\rev{supported}
tasks, which we call a {\em measurement box} or {\em box}.
Specifically, a task of \name can provide a set of task-specific parameters, which users \rev{can} specify
\rev{to instantiate tests}.
For instance, our memory task ($\S$\ref{design:micro}) allows users to specify which level of the DPU memory hierarchy they want to measure (i.e., L1--L3, or DRAM) and whether they measure throughput or latency. 
Users declare a box in a JSON file that specifies {\em the tasks to run}, {\em parameters for each task}, and {\em the metrics to measure}.
Figure~\ref{fig:box_example} shows a box example that specifies a job that measures 
(1) \rev{the median and tail latency and bandwidth for network communication with TCP on the DPU, using an increasing number of CPU cores;}
and (2) 
\rev{the throughput of scanning database tuples when pushing down the predicate to the DPU with specified core utilization.} 


\begin{figure}[t]
    \centering
    \begin{minipage}{.96\columnwidth}
    \begin{lstlisting} [language=python]
    "tasks":
    [
        {
            "task_name": "net_tcp",
            "parameters": {
                "data_size": [8,8192],
                "threads": [1,2,4,8],
                ..
            },
            "metrics": ["p50", "p99", "bandwidth"]
        },
        {
            "task_name": "pred_pushdown",
            "parameters": {
                "dpu_cores": [1,2,4]
            },
            "metrics": ["throughput"],
        },
    ]
    \end{lstlisting}
    \end{minipage}
    \caption{A box that includes \rev{a microbenchmark (network) and a cloud database module (predicate pushdown)}.}
    \label{fig:box_example}
    \vspace{-3mm}
\end{figure}

Moreover, \name allows users to 
\rev{create}
new tasks as {\em plugins}, a benefit of its extensible task abstraction.
Users can implement plugins for accessing vendor-specific DPU features such as hardware accelerators. 
\rev{To add a plugin to \name, a user can create a dedicated directory in \name's repository, under which she instantiates the task abstraction with four respective Python scripts. These scripts are the shells of arbitrary performance test implementations (i.e., in arbitrary language with arbitrary dependencies).}
Once a plugin is added, it can be included in a box. 
\rev{
The built-in tasks in \name are portable to any DPU SoC that offers standard Linux distribution and toolchain, which is the case for nearly every DPU we know.
Plugins, on the other hand, rely on special hardware and/or proprietary software, and thus portability is not expected.}





\subsection{Workflow}

\begin{figure}[t]
    \centering
        \centering
        \includegraphics[width=\columnwidth]{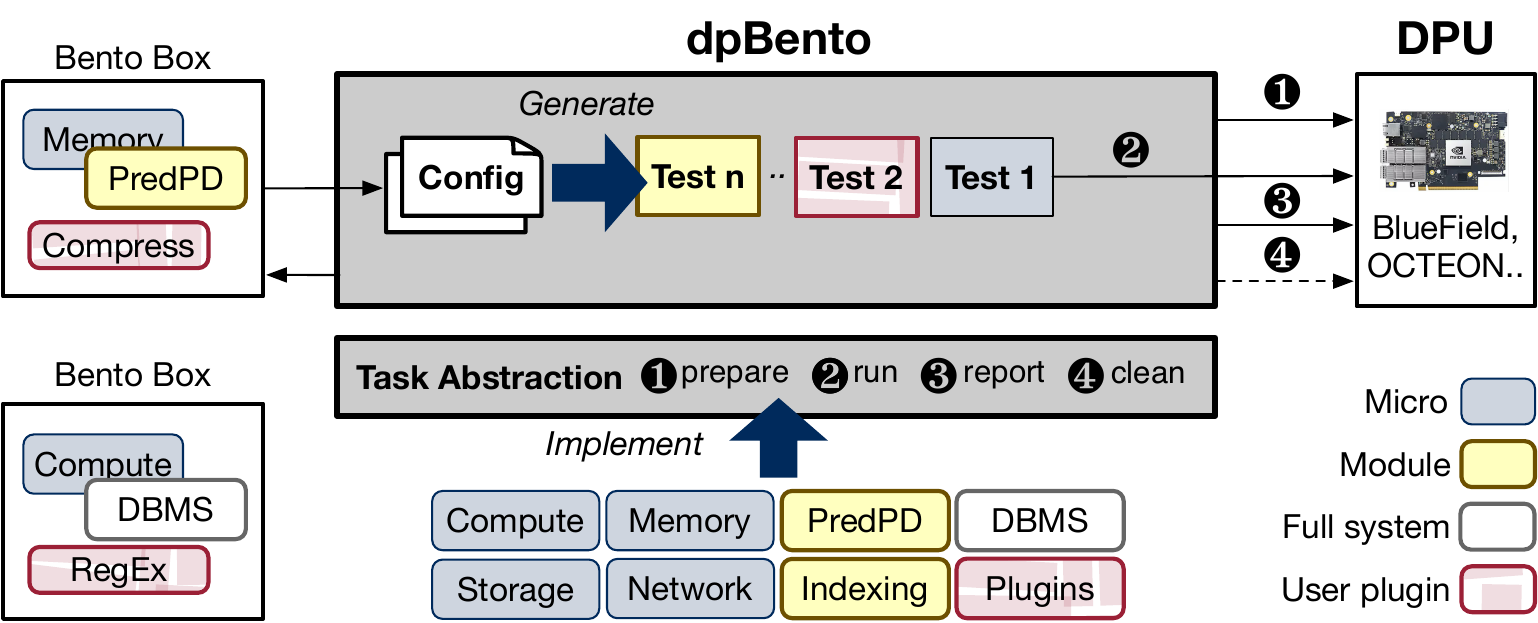}
        \caption{\name overview.}
        \label{fig:overview}
\end{figure}

Figure~\ref{fig:overview} shows the workflow of \name. 
It first parses the JSON file of each box to identify all tasks to be executed in that job.
For each task, it performs cross-product joins between parameters to generate all possible combinations, i.e., tests.
We do not join parameters and metrics as one test may produce results for multiple metrics, e.g., one \rev{network} test can generate both median and 99-percentile latencies.
Before executing any test, \ding{182} \name invokes the {\tt prepare} script to initialize the DPU environment for all tests in the current task to avoid repeated preparations between tests.
Then, \ding{183} the parameter combinations are sequentially passed, along with the metrics, to the {\tt run} script to perform a test.
Tests can individually generate logs to cache intermediate measurement results.
After all tests are executed, \ding{184} the {\tt report} script is invoked to process intermediate results and produce the final report to the user.

Performance testing is fully automated in \name.
{\em Users just need to declare boxes}, and all the above steps are transparently managed by the framework.
As multiple boxes may involve the same tasks, and preparation can be time-consuming, we do not invoke the {\tt clean} script immediately after each task.
Instead, \ding{185} a command line is provided for users to explicitly clean up the DPU.


\subsection{Microbenchmarks}
\label{design:micro}

We have implemented a range of data processing workloads as the built-in tasks in \name using its task abstraction, as shown in Table~\ref{tab:built-in}.
We first present microbenchmarks that use primitive operations to evaluate the compute and I/O performance of DPUs.

\subsubsection{Compute}
The CPU microbenchmark focuses on basic operations over primitive types, specifically, arithmetic over 
\rev{primitive numeric types}, which are heavily involved in data processing, e.g., for evaluating predicates and expressions, \rev{as well as string functions such as comparison, searching and manipulation.} 
\rev{Although decimal types are more often used than floating point in database systems, typical CPUs today have no hardware support for it, but do for integers and floating point.
Usage of decimal types usually relies on libraries or language features that have varying levels of optimized performance that do not directly reflect the hardware capability; additionally, fixed-point decimal type is commonly represented as an integer scaled by a power of ten \cite{decimal-perf}.
With the rise of vector databases, we expect floating point to be more heavily used by those systems due to the similarity search algorithms involved, such as \cite{diskann, HNSW, ivfoadc}.
In conjunction with quantization techniques \cite{QP}, computing with smaller data types (8, 16 bits etc.) becomes much more important. We therefore benchmark integers and floating-point numbers of different widths on DPUs.}
We evaluate the throughput (in operations per second) of arithmetic operations: addition, subtraction, multiplication, and division, on a single physical DPU core.
This task stresses the raw computing power by repeatedly performing the corresponding instructions over registers, ruling out the effect of the CPU cache and main memory.
Users can parameterize which data type (e.g., {\tt int8}, {\tt int128}, or {\tt float64}) and which operation ({\tt add}, {\tt sub}, {\tt mul}, or {\tt div}) to test.
\rev{String operations such as comparisons and substring matching are common in query predicates and directly affect scan and filter performance. We benchmark them over small (10\,B), medium (64\,B and 256\,B) and large (1\,KB) strings to mimic different application scenarios, and measure performance in operations per second.}


Besides the built-in CPU tests, 
we have also implemented a list of plugins, which are higher-level compute-intensive tasks, to assess the performance benefits of hardware accelerators on DPUs.
To differentiate from primitive type tests, we term these plugins as {\em optimizable tasks}---ones that can additionally be optimized with vectorized (SIMD) instructions, with multi-core parallelism, and with DPU hardware accelerators. 
We have included three such tasks: compression with the DEFLATE algorithm~\cite{deflate-algo}, correspondingly decompression, 
and regular expression (RegEx) matching.
These tasks are frequently invoked on the data path of cloud data systems~\cite{fpga_compenc}.
Exploring optimizable tasks gives us a glimpse into the performance gains with different levels of possible compute optimizations, especially the hardware accelerators that are yet to be fully exploited by today's data systems.

\begin{table}[t]
\small
\setlength{\tabcolsep}{1pt}
    \caption{\rev{Built-in tasks in \name.}}
    \centering\begin{tabular}{c c c}
        \toprule
        & {\bf Tasks} & {\bf Parameters} \\
    	\midrule
    	\multirow{4}{*}{Micro} & Compute & Data type, Operation \\
    	 & Memory & Operation, Object size, Pattern, \#Threads \\
         & Storage & I/O type, Access, Pattern, Depth, \#Threads \\
    	 & Network & Message size, Depth, \#Threads \\
        \midrule
    	 \multirow{2}{*}{\rev{Module}} & \rev{Pred pushdown} & \rev{Scale, Selectivity, \#Threads} \\
         & \rev{Index offloading}  & \rev{Scale, Op, Pattern, Split ratio, \#Threads} \\
        \midrule
    	 \multirow{1}{*}{Full system} & DBMS & Scale, Query, Execution mode, \#Threads \\
        \bottomrule
    \end{tabular}
	\label{tab:built-in}
\end{table}

\subsubsection{Memory}
Our memory microbenchmark task evaluates the speed of accessing in-memory objects. 
Specifically, we create a memory buffer of specified size, then access (read or write) pointer-size data within the buffer based on specified access pattern (either random access or sequential access), and finally measure the throughput (accesses completed per second) 
or bandwidth (GB/s).
In addition to these parameters, users can also configure the number of threads that issue memory accesses in parallel to better utilize memory bandwidth. 
We adopt sysbench~\cite{sysbench} as the test driver, which is a multi-threaded benchmark tool that can be scriptized.
When performing a DPU benchmark, the parameters declared for this task will be passed to sysbench to generate memory tests.




\subsubsection{Storage}
The storage microbenchmark generates asynchronous disk I/O activities with an extensive set of parameters to measure DPU-local storage device performance.
At the core, 
\rev{it is an extensive storage testing toolkit. Specifically, it initializes a file with random content on a specified disk, allows issuing file reads and writes, and measures detailed latency and throughput statistics of these operations.
To perform efficient secondary storage access, it enables asynchronous I/O with {\tt io\_uring} or {\tt libaio}.
Our experiments show that our storage benchmark tool can saturate the bandwidth of the local disks of all tested DPUs, as well as that of the host's NVMe SSD.}
Users can specify I/O type (reads or writes), access granularity (in bytes), queue depth (the number of outstanding file requests), and concurrency (the number of threads issuing I/O in parallel).
They can also select metrics of interest from throughput, average latency, and tail latency of different percentiles.




\subsubsection{Networking}
To evaluate network transfer performance, \name includes a benchmark task that measures the speed of TCP, the most popular transport in data systems, using Linux sockets.
Specifically, the task automatically creates two TCP endpoints on the source, which issues messages of specified size, and the destination, which receives each message and bounces it back to the source.
The source and destination can be the DPU or a host with a user-provided IP address (either the local host or a remote machine).
Our task measures the end-to-end latency and throughput of the message transmissions. 
Users can parameterize message size in bytes, queue depth (the number of outstanding messages in a TCP connection), and concurrency (the number of TCP connections, each processed by a client/server thread).

\subsection{\rev{Cloud Database Modules}}
\label{design:module}
\rev{Prior work has demonstrated two general advantages of offloading cloud database components to DPUs~\cite{dds, smartshuffle, ThostrupFZB22}: near-data processing, which executes operations closer to data to minimize data movement overhead, and augmented processing capabilities in addition to the host. We include two tasks in \name that offload cloud database modules to explore these benefits respectively.}

\subsubsection{\rev{Predicate Pushdown}}
\rev{This task targets a disaggregated storage architecture where SQL queries are executed by the database engine on the compute server, and database files are stored on a storage server.
The storage server is equipped with a DPU, which the compute server connects to with a 100\,Gbps link and we exploit for pushing down predicates when scanning database tuples. In the baseline, we run DuckDB (the DBMS detailed in $\S$\ref{design:macro}) on the compute server, and the database files are fetched via the network from the storage server to execute the scan. In the pushdown version, we execute the scan module of the DBMS on the DPU and return only qualified tuples to the compute server. We scan the {\tt lineitem} table in the TPC-H benchmark.}

\rev{As shown in Table~\ref{tab:built-in}, users can configure the scale of the table, the selectivity of the predicate, and how many threads are spawned on the DPU. 
The metric is tuples scanned per second.}

\subsubsection{\rev{Index Offloading}}
\rev{The second module we offload to the DPU is index. Specifically, we use the DPU equipped on a database server as a coprocessor of the host to partially serve index requests. We range-partition a B+ tree between the host and the DPU such that serving requests from the DPU can boost the overall index performance.
\pb{The implementation is adapted from LMDB~\cite{lmdb}, which supports concurrency with MVCC.}
We use the YCSB benchmark as the workload. A user of \name can configure the index size (KV record size and count), operation (read/write), access pattern (uniform or skewed), the ratio of the index offloaded, as well as the number of DPU threads executing the offloaded requests. This task measures index throughput (completed operations per second).}

\subsection{\rev{A Full DBMS}}
\label{design:macro}
To enable system-level performance testing \rev{and quantify the performance degradation of running an end-to-end system on low-power DPU cores, we include a full system evaluation task with DuckDB~\cite{duckdb}, a full-fledged \pb{relational DBMS that is light-weight, thus more suitable for DPUs}.}  
DuckDB targets embedded data analytics and optimizes deployment simplicity by avoiding any external dependencies.
The whole system can be compiled to a header file and a library that can be easily integrated into applications.
Running inside the application process, it minimizes data movement between the application and the DBMS.

Our task pulls the latest version of DuckDB (currently v1.1.\rev{3}) and compiles it from the source code in the {\tt prepare} stage.
It sets up TPC-H as the workload based on the specified scale factor(s).
Users can also parameterize which TPC-H queries to run, the number of threads created in DuckDB, and the execution mode (cold, where the queries are never executed on the DPU, or hot, where the queries have been executed a few times to warm up memory buffers). 
We focus on query execution time as the performance metric.

\section{Hardware to Benchmark}
\label{sec:hardware}
With \name, we have benchmarked several recent DPUs that are in mass production.
This section describes the hardware setups of the DPUs, as well as the host machine that we use as the baseline to compare to data processing performance on DPUs.

\heading{NVIDIA BlueField.}
We assess two generations of NVIDIA BlueField DPUs: BlueFiled-2 (BF-2) and its successor BlueField-3 (BF-3).
Both are in mass production~\cite{bf2_perf_char, bf2_rdma, dds}.
Figure~\ref{fig:dpu-arch} shows the details of the two DPUs in our testbed.
On BF-2, there is an Arm A72 CPU (64-bit), which consists of 8 cores @ 2.5\,GHz.
Every two cores share a 1\,MB L2 cache, and a 6\,MB L3 cache is shared between all cores.
There is 16\,GB onboard DDR\,4 memory that serves as the DPU's main memory.
It has a variety of hardware accelerators, including these for compression, decompression, and RegEx matching.
A ConnectX-6 NIC provides 100\,Gbps network bandwidth, and the board is connected to the host via PCIe 4.0.
BF-2 also has an eMMC flash device that is soldered directly onto its motherboard.

BF-3 is an upgrade from BF-2 in nearly every aspect: the Arm CPU is upgraded to A78, which has 16 cores with clock rates up to 3.0\,GHz.
L2 cache increases to 6\,MB and L3 to 16\,MB.
The onboard memory levels up to DDR\,5, 32\,GB.
The network interface is upgraded to ConnectX-7, providing 400\,Gbps bandwidth, and the interface to the host is advanced to PCIe 5.0.
The directly attached local storage is also updated to a 160\,GB NVMe SSD.
Interestingly, the compression engine is removed, which makes a different set of hardware accelerators.

\heading{Marvell OCTEON.}
Also shown in Figure~\ref{fig:dpu-arch} is a Marvell OCTEON TX2 DPU in our testbed.
Same as BF-2, it is equipped with an Arm A72 CPU but with more cores (24 cores @ 2.2\,GHz), with a 1\,MB L2 cache shared between two cores and another 14\,MB L3 cache shared across all cores.
The DPU has 32\,GB onboard DDR\,4 memory, and provides a 100\,Gbps Ethernet network interface and PCIe 3.0 for connecting to the host.
The hardware accelerators are different from BlueField, primarily for network security and packet processing.
Like \nvbf, our \octeon has an eMMC local storage of 64\,GB.

\heading{Host Machine.}
We conduct the same set of \name tasks on a host server, which was recently assembled, and thus represents the state-of-the-art host configuration.
Specifically, the server has two AMD EPYC 9254 24-Core CPUs @ 2.9\,GHz, 
so in total there are 48 cores (96 hyperthreads), 48\,MB L2 cache, and 256\,MB L3 cache.
The main memory on the server is 128\,GB DDR\,5, 
and for persistent storage, there are two 960\,GB NVMe SSDs.
Comparing the benchmark results to those on the host helps better interpret data processing performance on the DPUs.

\section{Compute Efficiency}
\label{sec:compute-efficiency}

We first present the benchmark results of the compute and memory tasks on the three DPUs and the host.
Detailed findings follow.

\begin{figure*}
    \centering
    \begin{subfigure}{.33\textwidth}
        \centering
        \includegraphics[width=\textwidth]{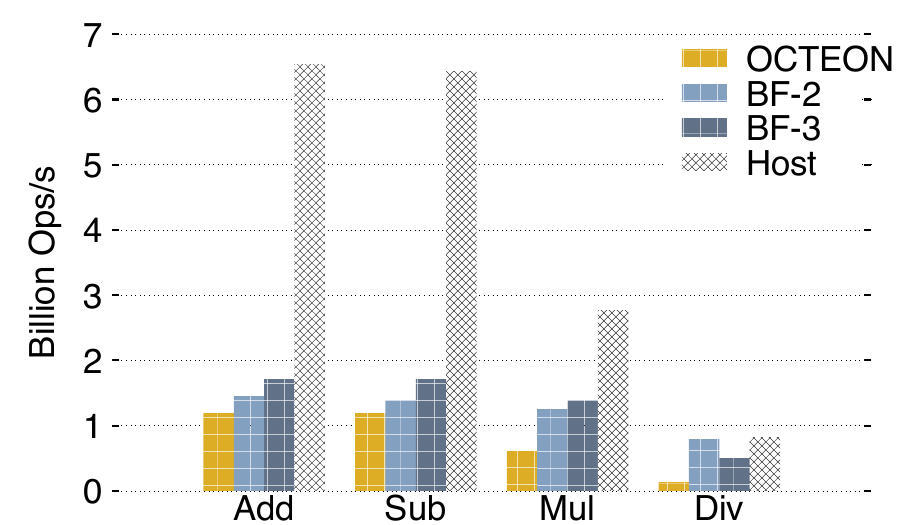}
        \caption{Results on {\tt int8}.}
        \label{fig:compute_int8}
    \end{subfigure}
    \begin{subfigure}{.33\textwidth}
        \centering
        \includegraphics[width=\textwidth]{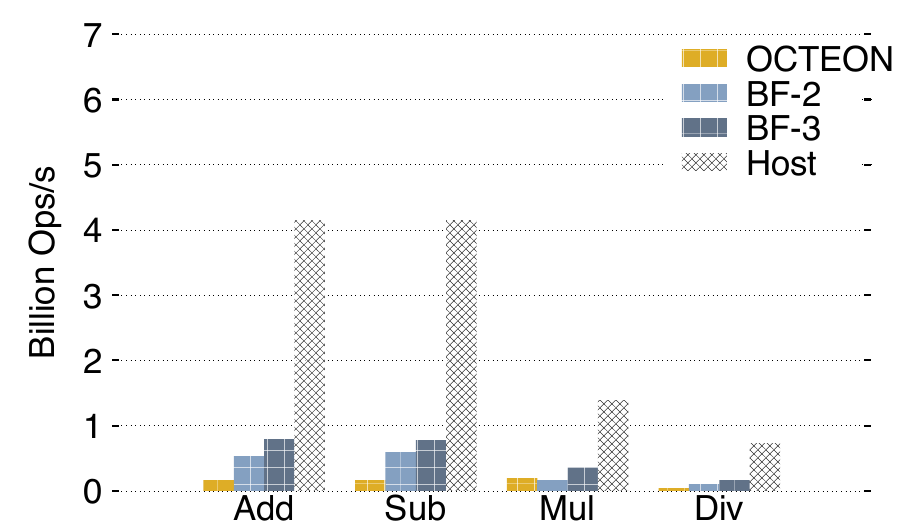}
        \caption{Results on {\tt int128}.}
        \label{fig:compute_int128}
    \end{subfigure}
    \begin{subfigure}{.33\textwidth}
        \centering
        \includegraphics[width=\textwidth]{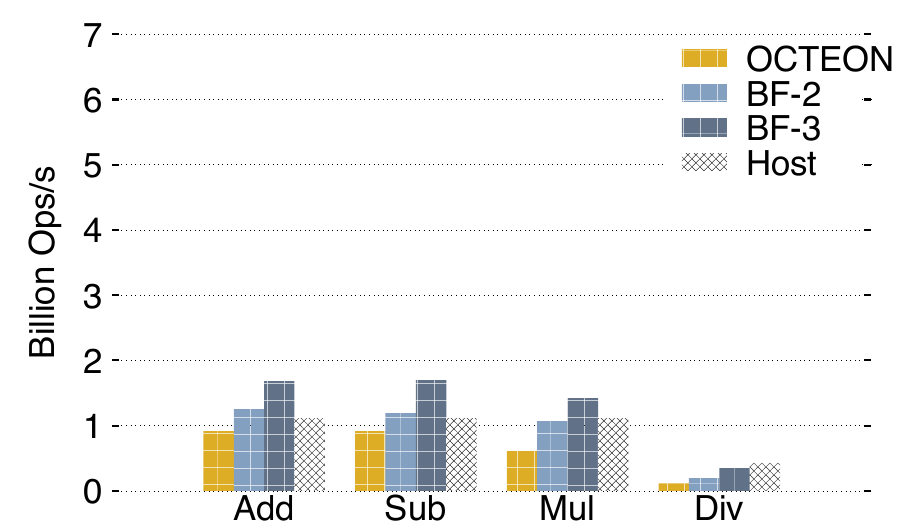}
        \caption{Results on {\tt fp64}.}
        \label{fig:compute_fp64}
    \end{subfigure}
    \caption{Benchmarking DPUs with primitive arithmetic operations on integers and floating-point numbers.}
    \label{fig:compute}
\end{figure*}

\begin{figure*}
    \centering
    \begin{subfigure}{.33\textwidth}
        \centering
        \includegraphics[width=\textwidth]{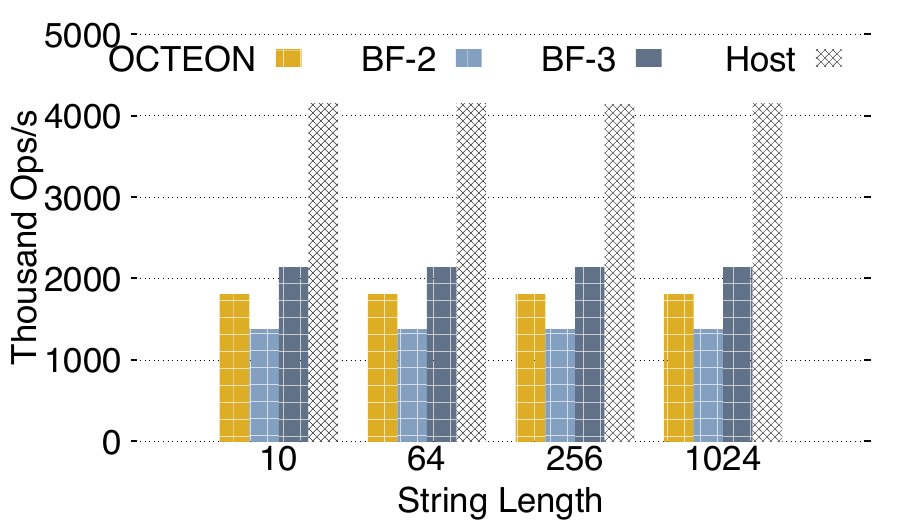}
        \caption{Results on {\tt strcmp}.}
        \label{fig:compute_strcmp}
    \end{subfigure}
    \begin{subfigure}{.33\textwidth}
        \centering
        \includegraphics[width=\textwidth]{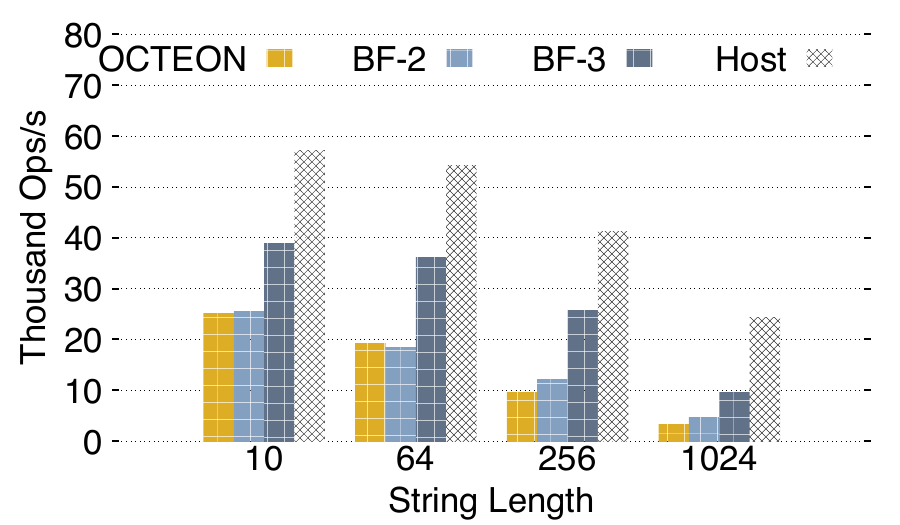}
        \caption{Results on {\tt strcat}.}
        \label{fig:compute_strcat}
    \end{subfigure}
    \begin{subfigure}{.33\textwidth}
        \centering
        \includegraphics[width=\textwidth]{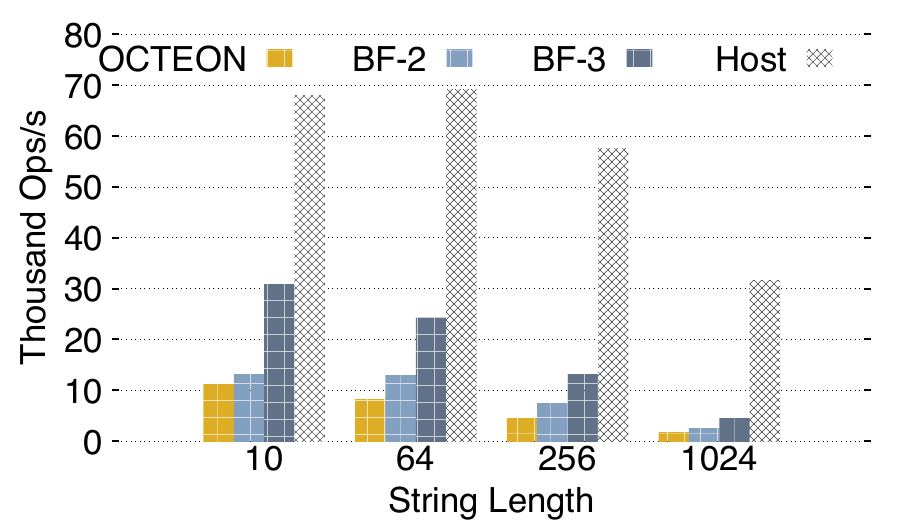}
        \caption{Results on {\tt strxfrm}.}
        \label{fig:compute_xfrm}
    \end{subfigure}
    \caption{Benchmarking DPUs with primitive string operations.}
    \label{fig:string}
\end{figure*}

\subsection{Primitive Operations}
\label{sec:comp-primitive}
Table below shows the testing parameters that we use to benchmark DPUs with the primitive compute task: we run all arithmetic operations on integers and floating-points of different sizes ({\tt int8}, {\tt fp64}, and {\tt int128}). 
These types and operations are commonly seen in different data systems, e.g., databases and ML models.

\begin{table}[h]
\small
\setlength{\tabcolsep}{0pt}
    \centering
    \begin{tabular}{c c c}
        \toprule
        & {\bf Data Type} & {\bf Operation} \\
        \midrule
        {\bf Arithmetic} & {\tt int8}, 
        {\tt fp64},
        {\tt int128} 
         & {\tt add}, {\tt sub}, {\tt mul}, {\tt div}\\
        \rev{{\bf String}} & \rev{{\tt str10}, {\tt str64},
        {\tt str256},
        {\tt str1024}} 
         & \rev{{\tt cmp}, {\tt cat}}, {\tt xfrm}\\
        \bottomrule
    \end{tabular}

    \label{tab:CPU-bench-setup}
\end{table}

The detailed benchmark results are shown in Figure~\ref{fig:compute} for both integers and floating-point numbers.
First, Figure~\ref{fig:compute_int8} shows the performance of the 8-bit integer operations.
For {\tt add} and {\tt sub}, the host AMD CPU demonstrates superior performance advantages compared to the DPU Arm cores, achieving 6.5 billion operations per second (ops/s)---up to 5.5$\times$ higher than the DPUs. 
For more complex {\tt mul} and {\tt div}, although all four CPUs exhibit a throughput drop as expected, the degree of degradation is different for the host CPU compared to the DPUs' CPUs.
Specifically, the host CPU experiences a 58\% throughput decrease when performing {\tt mul} compared to {\tt add} on {\tt int8}, higher than the degradation seen on OCTEON (49\%) and much higher than that of BF-2 (14\%) and BF-3 (19\%).
However, the host CPU is still $2\times$ better than the best DPU (BF-3).
For {\tt div}, a further decrease in performance can be observed on the host CPU (70\% throughput drop compared to {\tt mul}).
The Arm cores on OCTEON follow the same trend: a significant 80\% throughput decrease. 
BF-2 and BF-3 perform relatively better, with lower throughput degradation from {\tt mul} (36\% and 64\%, respectively).

With {\tt int128}, as shown in Figure~\ref{fig:compute_int128}, we observe similar trends in performance across the board---the host is still the best performer by a large margin, and all CPUs see their throughput drop in {\tt mul} and {\tt div} compared to {\tt add} and {\tt sub}.
With increased operand sizes, throughputs of all operations on all platforms decrease accordingly, with DPUs seeing larger percentage drops across the board.
Specifically, from {\tt int8} to {\tt int128}, the host experiences 34\% throughput decrease on average across the operations, while the decrease is 76\%, 73\%, and 63\% for \octeon, \nvbf, and \nvbff, respectively.
The difference is especially significant in {\tt mul} and {\tt div}, e.g., the throughput decrease of the DPUs is 63\%--77\%, while it is only 12\% for the host, which is now 4.7$\times$ faster than the best-performing DPU (it was 1.7$\times$ on {\tt int8}).
These results show that the DPUs are better at handling smaller operands.


Floating point ({\tt fp64}) performance paints a very different picture for the DPUs, as seen in Figure ~\ref{fig:compute_fp64}.
The highly noticeable performance lead of the host CPU in integer compute is almost completely lost, where the two BlueField DPUs now outperform the host CPU in {\tt add}, {\tt sub}, and {\tt mul}, where \nvbff leads by more than 50\% on average.
\octeon also becomes relatively much more competitive with floating point compute, compared to its integer performance, albeit still trailing behind the BlueField family of DPUs and the host CPU by a fair margin.
It is worth noting that the host CPU still holds an advantage with division operations, although the lead is much smaller than that of integers.
The high performance of DPUs on floating-point operations can be ascribed to the hardware support in the Arm architecture~\cite{arm_float}.

\rev{
We have benchmarked three representative string operations: comparison ({\tt strcmp}), simple manipulation ({\tt strcat}), and complex transformation ({\tt strcfrm}), and Figure ~\ref{fig:string} shows the results. 
The host CPU has higher throughput in all categories, but the performance gap between the host CPU and DPUs varies across categories.
For string comparison, string size matters little for DPUs and the host CPU, where the host CPU has close to 2$\times$ the performance of the most powerful DPU (BF-3).
For string manipulation, the host still has the absolute performance advantage, but the lead over DPUs are smaller, especially for small string sizes: BF-3 achieves 68\% the performance of the host CPU, with 10\,B strings, dropping to 39\% for 1024\,B strings.
String transformation operations paint a different picture---the gap widens in favor of the host CPU as the string size increases, and that BF-3 holds a more substantial lead over the other DPUs than that of string search. However, the host CPU still has more than two times the performance of BF-3, where the gap widens with larger string sizes, reaching more than 7$\times$ the throughput on OCTEON.}

\begin{figure*}
    \centering
    \begin{subfigure}{.33\textwidth}
        \centering
        \includegraphics[width=\textwidth]{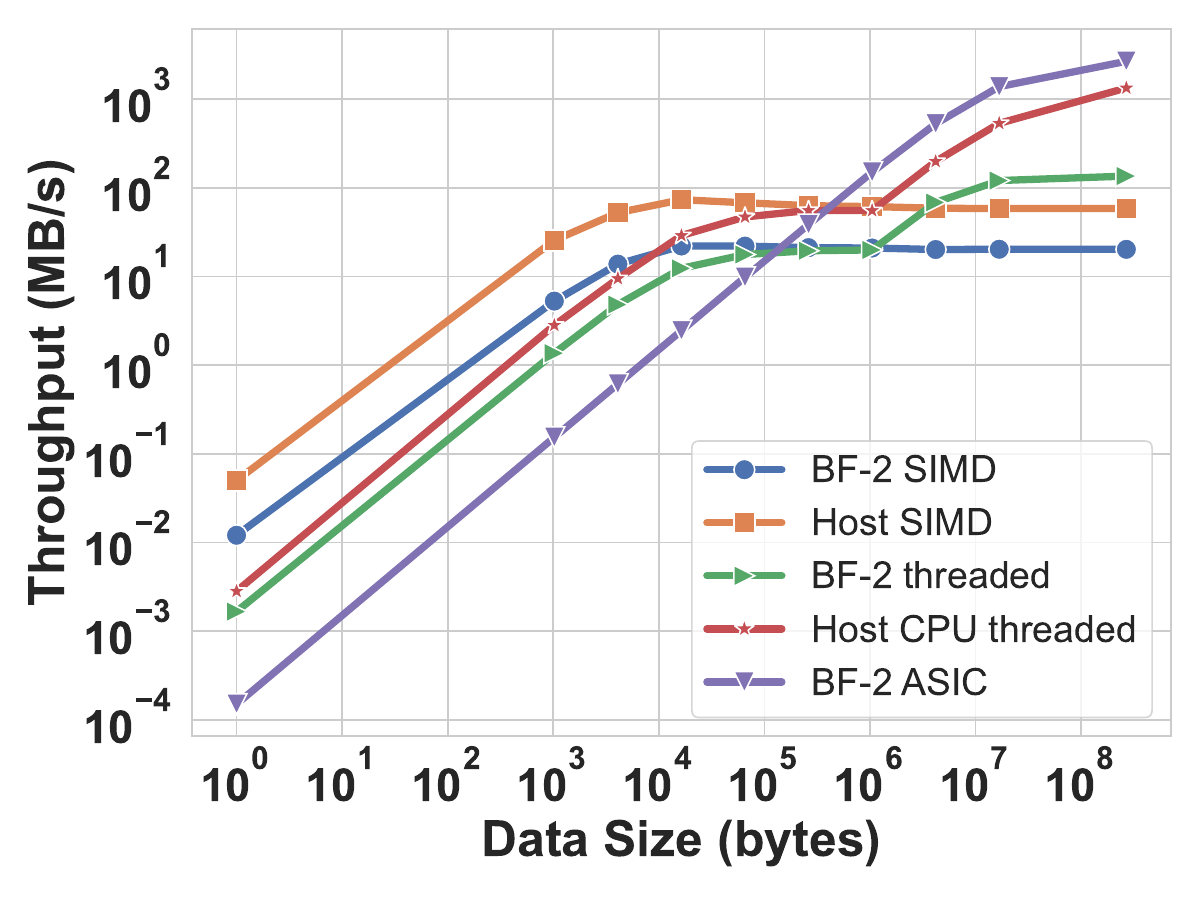}
        \caption{DEFLATE compression.}
        \label{fig:compression}
    \end{subfigure}
    \begin{subfigure}{.33\textwidth}
        \centering
        \includegraphics[width=\textwidth]{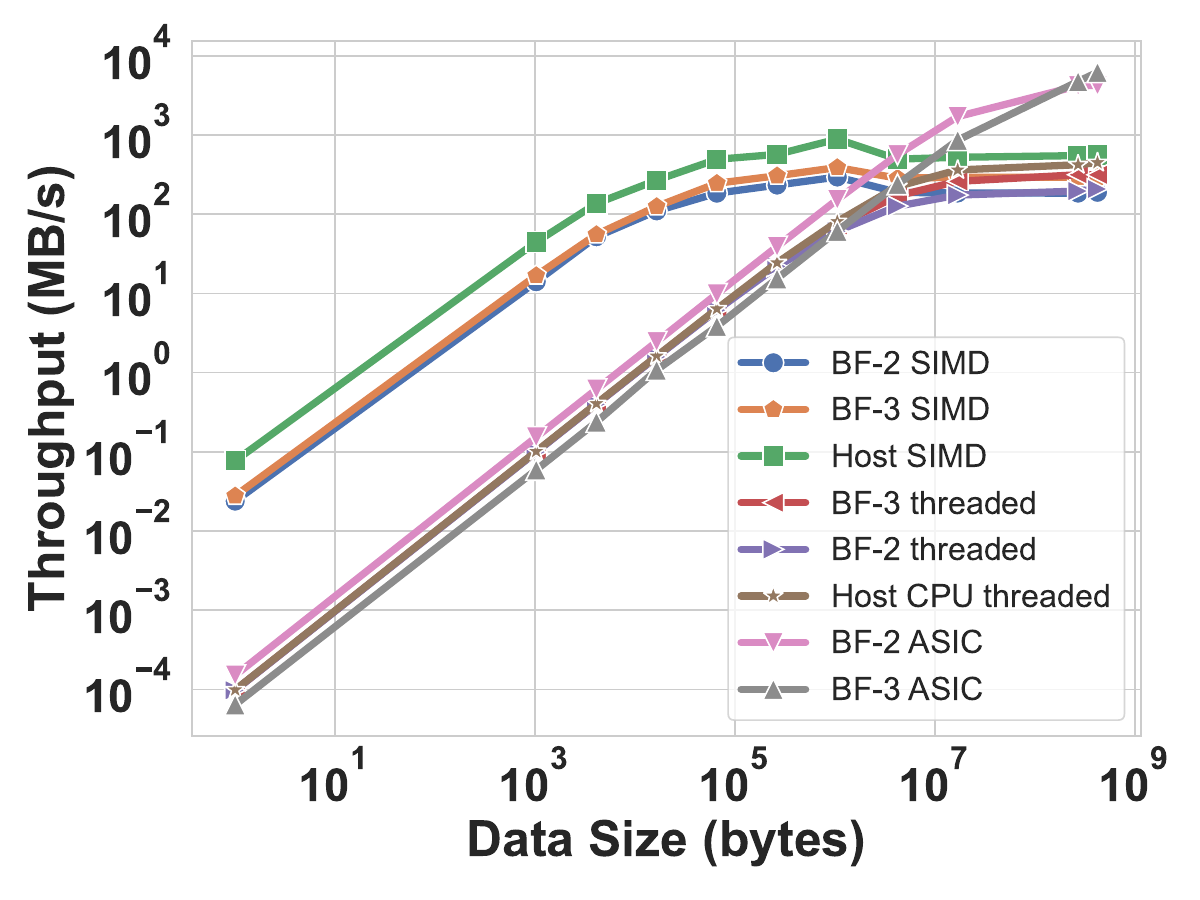}
        \caption{Decompression.}
        \label{fig:decompression}
    \end{subfigure}
    \begin{subfigure}{.33\textwidth}
        \centering
        \includegraphics[width=\textwidth]{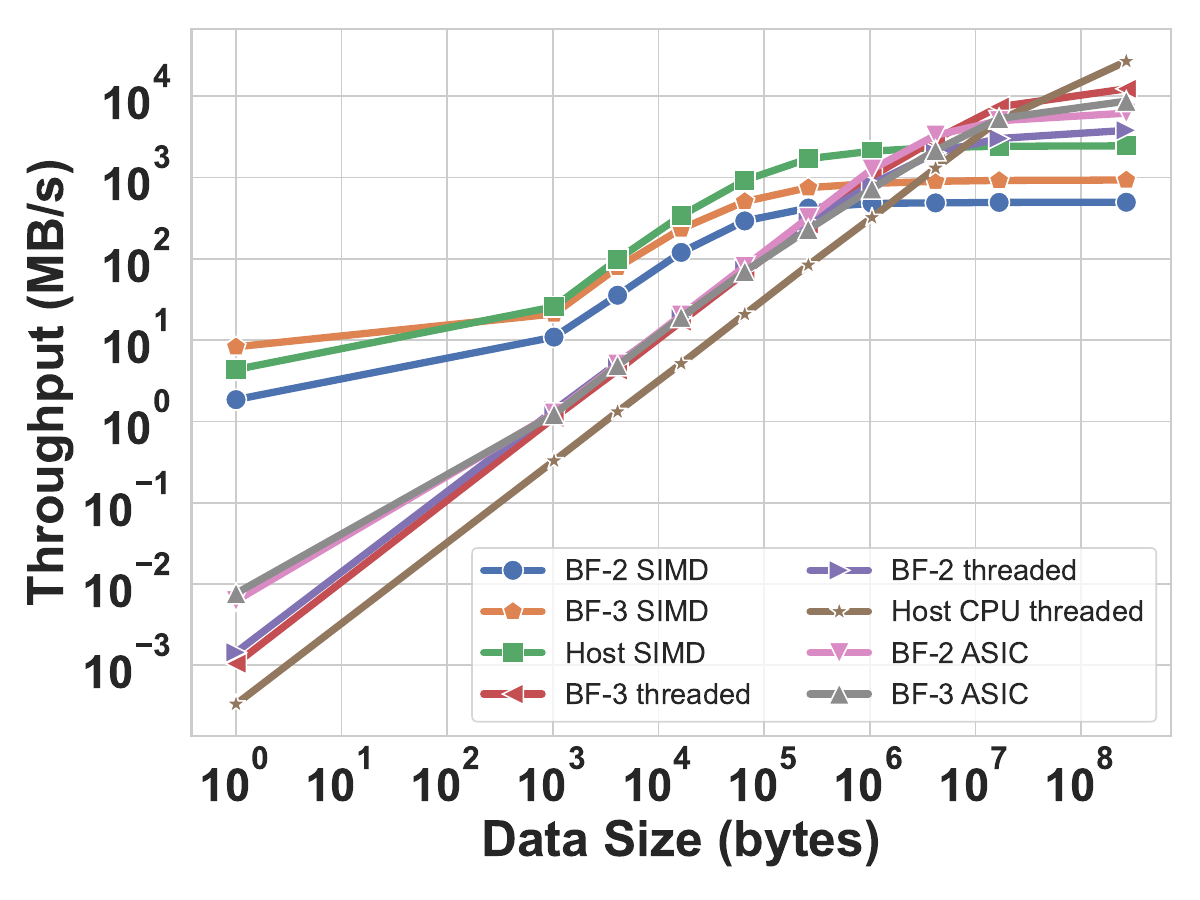}
        \caption{RegEx matching.}
        \label{fig:regex}
    \end{subfigure}
    \caption{Benchmarking DPUs with optimizable tasks that can be accelerated by hardware techniques.}
    \label{fig:compute_opt}
\end{figure*}

Based on these results, we have the following findings.
\vspace{3mm}
\begin{mdframed}[style=FindingFrame,nobreak=false,align=center,userdefinedwidth=27em]

\begin{itemize}
    \item {\em DPUs are faster at processing smaller operands and {\bf\em can even outperform the host for floating-point processing}.}
    \rev{\item {\em For strings, DPUs are more suitable for simpler operations.}}
\end{itemize}

\end{mdframed}

\subsection{Hardware Acceleration}
\label{sec:comp-accel}

To investigate the benefits of DPU hardware acceleration, and compare the performance characteristics with that of existing CPU techniques, we run three plugin tasks for NVIDIA DPUs.
In each task, we use DOCA~\cite{doca} to access the hardware accelerator on \nvbf and/or \nvbff.
A software version that can leverage SIMD and multi-threading is also implemented for running on CPUs.

We first examine compression using DEFLATE~\cite{deflate-algo}, one of the most commonly used compression algorithms in database systems~\cite{compressioni-survey}.
Specifically, we use this algorithm to compress strings generated from TPC-H \verb|orders| table of specified size.
\nvbf offers a hardware accelerator for this compression.
In addition to single-core performance of \nvbf and the host, we also compare \nvbf hardware acceleration to multi-threading (with all available cores) and SIMD, two techniques that can boost the performance of these two CPUs.
Figure~\ref{fig:compression} shows the compression speeds of different hardware techniques.
We first observe that the DPU hardware acceleration is not always desirable: there is a fixed startup overhead when invoking the hardware accelerator on \nvbf.
For data below 100\,KB, the performance of hardware offloading is lower than that on host and \nvbf CPUs. 
However, when data size increases ($\geq$1\,MB), \nvbf hardware offloading shows superior performance, offering higher throughput than threaded execution on the host CPU (4.9$\times$ faster when compressing 512\,MB data).
We note that for very small data sizes, multi-threaded execution also provides no benefits due to threading overhead.

Decompression acceleration is supported by both \nvbf and \nvbff, and the benchmark results are shown in Figure ~\ref{fig:decompression}.
Similar to compression results, the hardware accelerators on the BlueFields incur a relatively high startup latency, which is reflected in low throughput for smaller payload sizes (lower than 1\,MB), but are much more efficient with larger sizes: \nvbf decompression accelerator is 13$\times$ / 21$\times$ faster than the host CPU / its onboard CPU with multi-threading when compressing 256\,MB data.
Additionally, the \nvbff's accelerator exhibits a higher startup latency than that on \nvbf, but becomes faster than its last-generation counterpart as the data size grows to 100s of\,MB.
Another observation is that for decompression, the performance gap between the host and onboard CPUs is relatively smaller, especially with threaded execution, this is likely due to the nature of DEFLATE algorithm, where decoding serializes data access and is thus hard to parallelize. 

\begin{figure*}
    \centering
    \begin{subfigure}{.245\textwidth}
        \centering
        \includegraphics[width=\textwidth]{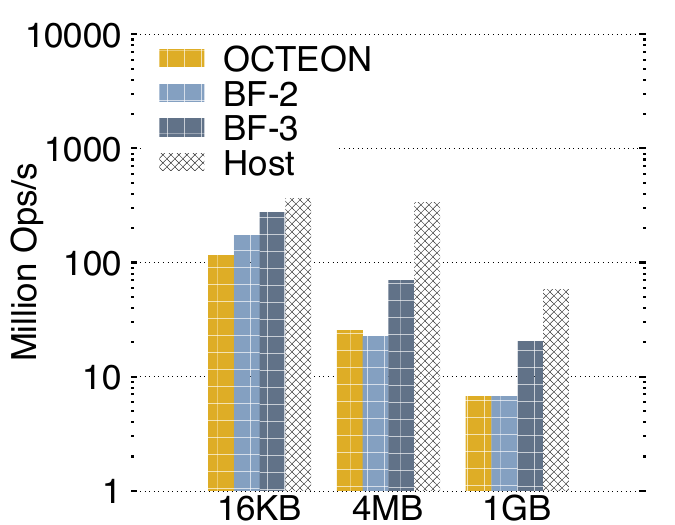}
        \caption{Random reads.}
        \label{fig:mem_read_rnd}
    \end{subfigure}
    \begin{subfigure}{.245\textwidth}
        \centering
        \includegraphics[width=\textwidth]{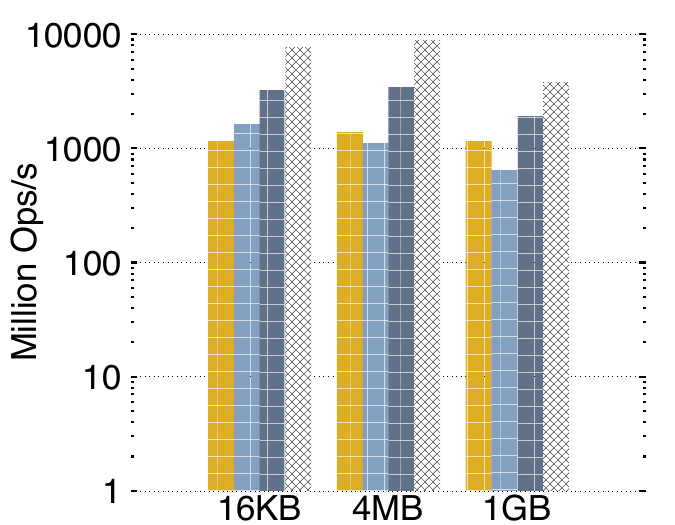}
        \caption{Sequential reads.}
        \label{fig:mem_read_seq}
    \end{subfigure}
    \begin{subfigure}{.245\textwidth}
        \centering
        \includegraphics[width=\textwidth]{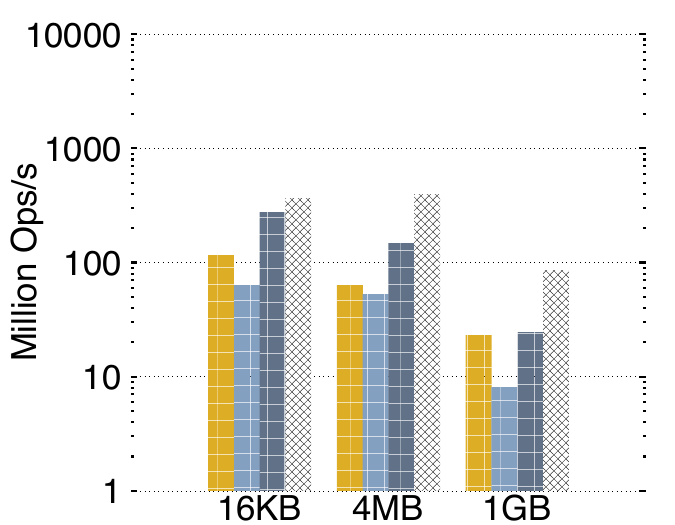}
        \caption{Random writes.}
        \label{fig:mem_write_rnd}
    \end{subfigure}
    \begin{subfigure}{.245\textwidth}
        \centering
        \includegraphics[width=\textwidth]{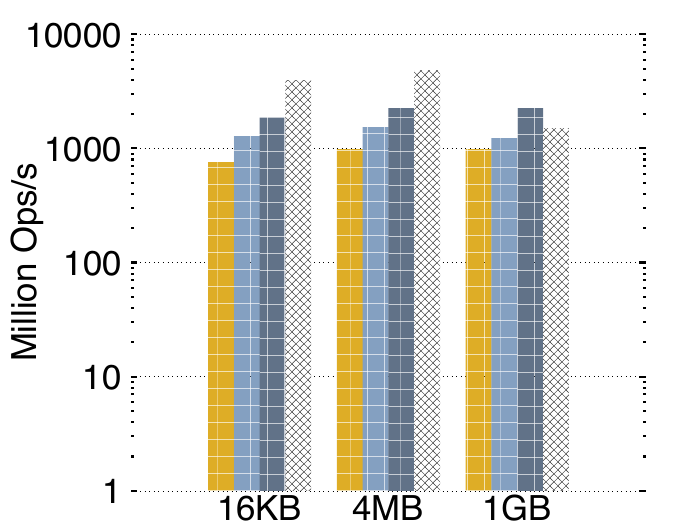}
        \caption{Sequential writes.}
        \label{fig:mem_write_seq}
    \end{subfigure}
    \caption{Benchmarking the memory efficiency of DPUs with varying object sizes.}
    \label{fig:mem_size}
\end{figure*}

Our final optimizable task is RegEx matching, where both \nvbf and \nvbff offer hardware acceleration.
In this task, we match the pattern that is part of TPC-H query 13, specifically, the generated pattern is \verb|like '%special%requests%'|.

Figure ~\ref{fig:regex} shows the result of varying the size of the string dataset.
First, the hardware-accelerated performance of \nvbf and \nvbff is identical, which is better than using threaded execution for small data sizes.
However, a single-threaded implementation with SIMD provides much better performance.
As data size increases, the hardware accelerators achieve comparable performance to multi-threaded execution (with all available CPU cores).
However, eventually the latter scales better: on 256\,MB data, the host CPU and \nvbff CPU are 3$\times$ and 1.4$\times$ faster than the RegEx accelerator on the two DPUs, respectively.
Using all cores for RegEx matching may not always be feasible or desirable---in that case, DPU hardware accelerators handily lead over the single-threaded SIMD counterpart and provides considerable CPU savings.


We make the findings below about hardware accelerators on DPUs from this experiment.
\vspace{3mm}
\begin{mdframed}[style=FindingFrame,nobreak=false,align=center,userdefinedwidth=27em]

\begin{itemize}
    \item {\em Hardware accelerators do not always outperform CPUs. Even when they do, {\bf\em they can improve throughput}, not latency.}
    \item {\bf\em Offloading compute-intensive tasks to hardware accelerators can save many CPU cycles.}
\end{itemize}

\end{mdframed}


\subsection{Memory}
\label{sec:mem}


The parameters that we use to benchmark memory performance of DPUs with the memory task are listed in 
the following table. 

\begin{table}[h]
\small
\setlength{\tabcolsep}{3pt}
    \centering
    \begin{tabular}{c c c c}
        \toprule
        {\bf Operation} & {\bf Object Size} & {\bf Pattern} & {\bf \#Threads} \\
        \midrule
        Read, Write &
        16\,KB, 4\,MB, 1\,GB &
        Random, Sequential &
        1--Max\\
        \bottomrule
    \end{tabular}
    \label{tab:mem-bench-setup}
\end{table}

We first use one thread to issue random and sequential reads and writes to all the above memory sizes. Figure~\ref{fig:mem_size} shows the throughput of pointer-size memory accesses.
When performing random reads to an object as small as 16\,KB, the object can be cached in L2 cache for all platforms, and thus the accesses are efficient.
As shown in Figure~\ref{fig:mem_read_rnd}, all the DPUs and the host can achieve higher than 100 million ops/s random access throughput.
Between DPUs, \nvbff achieves the highest throughput, 1.6$\times$ higher than \nvbf.
The gap is even larger than that between the host and \nvbff (1.3$\times$).
When we increase the object size to 4\,MB, the throughputs of the DPUs drop dramatically: 78\%, 87\%, and 75\% decrease for \octeon, \nvbf, and \nvbff, respectively.
At this size, the working set is very likely to spill to to L3 for the DPUs.
As the host has a much larger L2 cache (48\,MB), its throughput remains high.
To eliminate CPU caching effect, we further test a 1\,GB buffer, which exceeds the CPU caches to a large extent.
We can see that the throughput drops to the next level for the platforms: the throughput of the host drops to 58 million ops/s---a 83\% drop, followed by \nvbff, achieving 20 million op/s, and then by \octeon and \nvbf, both at 6.7 million ops/s.
This experiment shows that for random reads, DPUs are good at caching small object with comparable performance to the host.
When the gap to the host increases with the size of the buffered objects.

\begin{figure}[t]
    \centering
        \centering
        \includegraphics[width=\columnwidth]{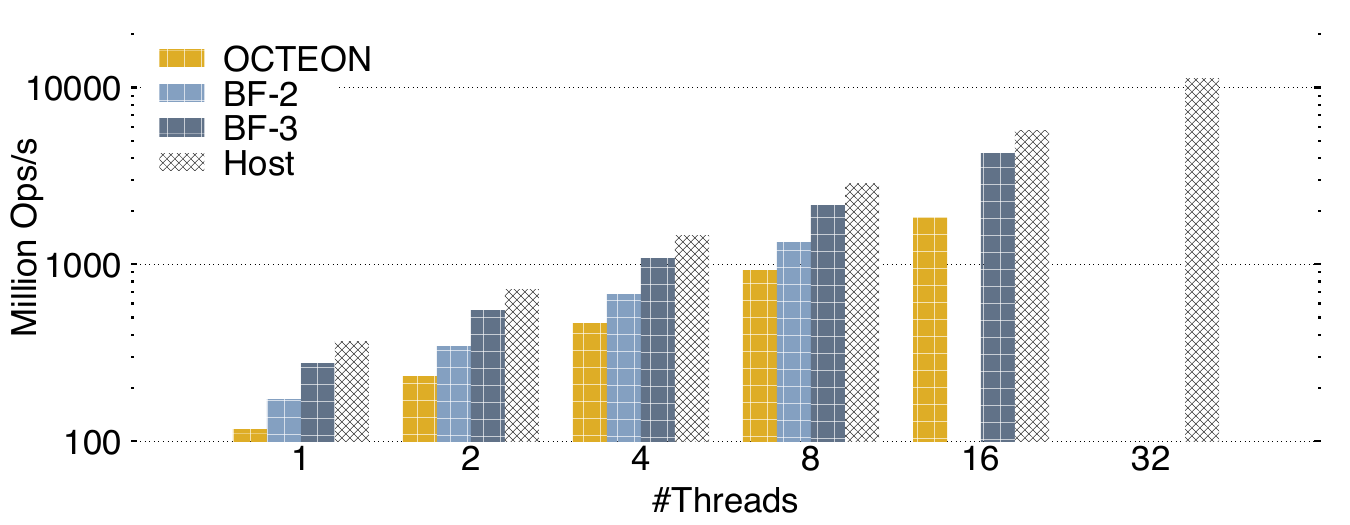}
        \caption{Scaling up memory accesses (\rev{random reads)}. \nvbf, \nvbff, and \octeon have 8, 16, and 24 cores, respectively.}
        \label{fig:mem-scale}
\end{figure}

\begin{figure*}
    \centering
    \begin{subfigure}{.245\textwidth}
        \centering
        \includegraphics[width=\textwidth]{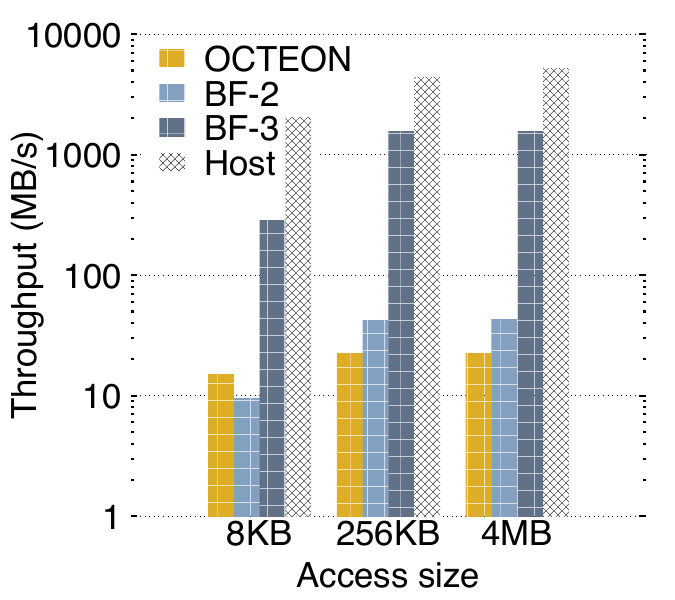}
        \caption{Random reads.}
        \label{fig:stor_read_rnd}
    \end{subfigure}
    \begin{subfigure}{.245\textwidth}
        \centering
        \includegraphics[width=\textwidth]{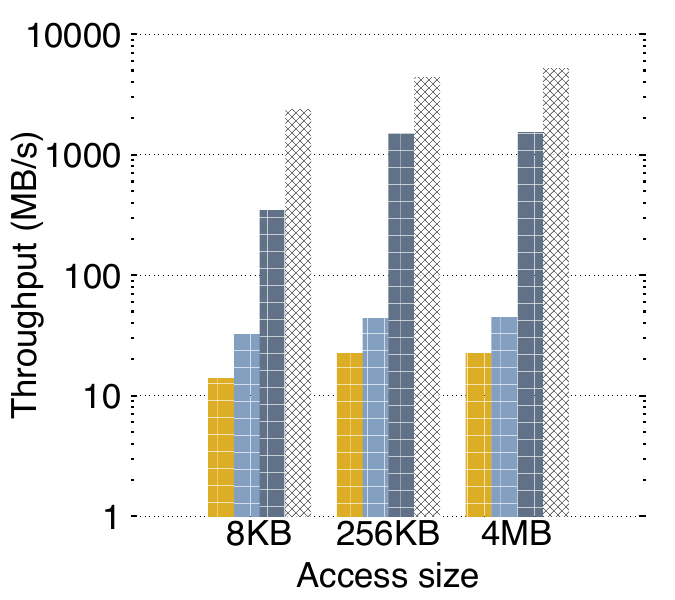}
        \caption{Sequential reads.}
        \label{fig:stor_read_seq}
    \end{subfigure}
    \begin{subfigure}{.245\textwidth}
        \centering
        \includegraphics[width=\textwidth]{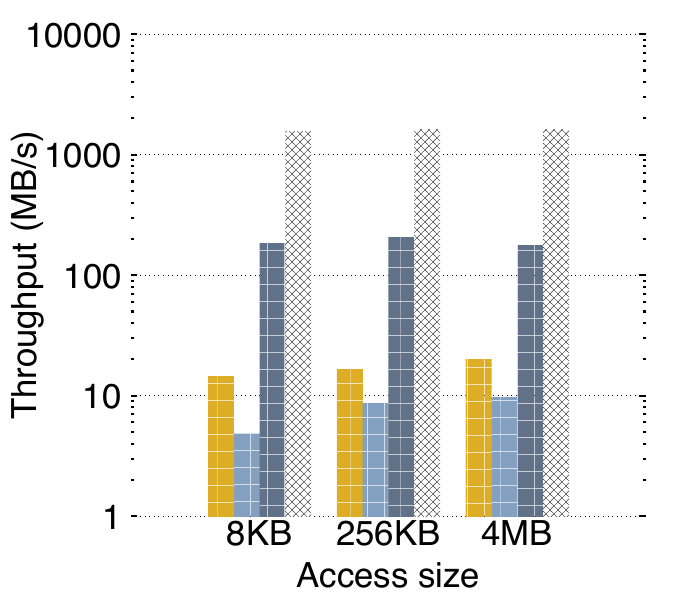}
        \caption{Random writes.}
        \label{fig:stor_write_rnd}
    \end{subfigure}
    \begin{subfigure}{.245\textwidth}
        \centering
        \includegraphics[width=\textwidth]{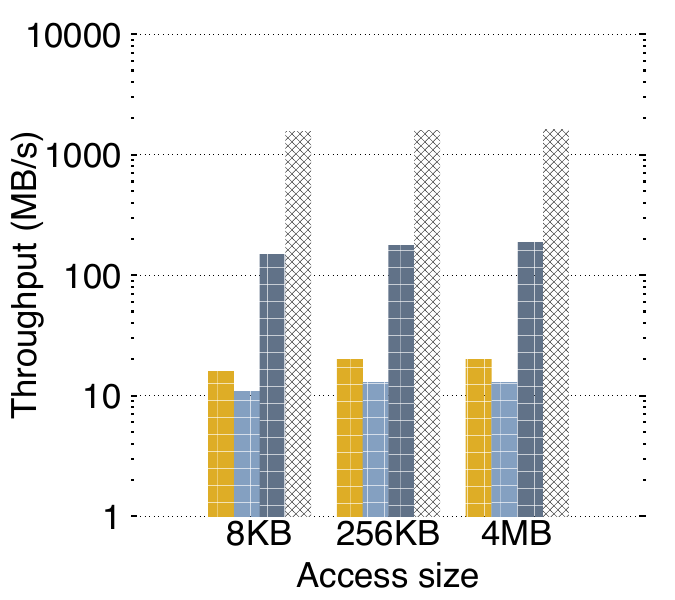}
        \caption{Sequential writes.}
        \label{fig:stor_write_seq}
    \end{subfigure}
    \caption{Benchmarking DPU and host local storage throughput. We vary the I/O access size from 8~KB to 4~MB, and evaluate the storage I/O throughput under random/sequential read and write patterns.}
    \label{fig:stor_thr}
\end{figure*}

The above findings can be applied to random writes as well (Figure~\ref{fig:mem_write_rnd}).
All platforms witness throughput drop when accessing objects larger than their L3 cache.
The best-performing DPU remains \nvbff, whose gap to the host enlarges as the memory buffer size increases.
However, \octeon now performs significantly better than \nvbf and approaches \nvbff for writing to a 1\,GB buffer.

In sequential accesses, CPU prefetching plays a critical role.
Figures~\ref{fig:mem_read_seq} and~\ref{fig:mem_write_seq} show the performance of sequentially reading and writing memory objects, which are much higher than random accesses, as expected.
We make the following observations.
First, prefetching on the DPUs is as effective as it is on the host: the throughputs of all platforms are largely stable when varying the object size from 16\,KB to 1\,GB for both reads and writes.
Second, the gap between the DPUs and the host is not as large as that when performing random accesses, especially for medium and large sizes, e.g., the host is now $5.9\times$ faster than \nvbf for sequential reads (vs. 8.6$\times$ for random reads).
Finally, DPUs can even achieve higher throughput than that of the host.
In particular, when sequentially writing to a 1\,GB memory, \nvbff achieves 2.2 billion ops/s, which is higher than the host throughput at 1.5 billion ops/s.

Figure~\ref{fig:mem-scale} demonstrates how CPU core count may affect memory accesses.
Specifically, multiple threads randomly access a 16\,KB memory buffer in parallel.
We can see that a single thread is incapable of saturating memory bandwidth, and the achieved throughput increases linearly with thread count.
However, the DPUs have fewer cores (8, 16, and 24 on \nvbf, \nvbff, and \octeon, respectively, vs. 96 cores on the host), which limits the their highest memory access performance: 1.3 billion op/s on \nvbf, 2.7 billion op/s on \octeon, and 4.3 billion op/s on \nvbff.
In comparison, the host achieves 11.3 billion op/s with 32 cores.

Memory benchmark results are summarized as follows.

\vspace{3mm}
\begin{mdframed}[style=FindingFrame,nobreak=false,align=center,userdefinedwidth=27em]

\begin{itemize}
    \item {\em DPUs are good at sequentially accessing in-memory objects. Sometimes, {\bf\em they even outperform the host}.}
    \item {\em If the memory accesses are random, DPUs are better at accessing small objects than the larger ones.}
    \item {\em DPUs' limited core count can become a bottleneck for high-throughput memory access.}
\end{itemize}

\end{mdframed}
\section{I/O Efficiency}
\label{sec:io-efficiency}

Next, we evaluate the I/O performance, including both storage I/O and networking I/O, of the three DPUs.
We also use the host I/O performance as a baseline.

\subsection{Storage}
\label{sec:stor}
The testing parameters for our storage benchmarking are shown in the following table. 
Like memory benchmarking, we evaluate the performance of both sequential and random reads and writes.
We vary access granularity, the number of outstanding requests, and parallelism.
Since storage I/O is asynchronous, we measure both latency and throughput.

\begin{table} [h]
\small
\setlength{\tabcolsep}{2pt}
    \centering
    \begin{tabular}{c c c c c}
        \toprule
        {\bf I/O Type} & {\bf Access Size} & {\bf Pattern} & {\bf Queue Depth}  & {\bf \#Threads} \\
        \midrule
        Read, Write &
        8\,KB--4\,MB &
        Random, Sequential &
        1--256 &
        1--Max\\
        \bottomrule
    \end{tabular}
    \label{tab:storage-bench-setup}
\end{table}

\begin{figure}[t]
    \centering
    \begin{subfigure}{\columnwidth}
        \centering
        \includegraphics[width=\textwidth]{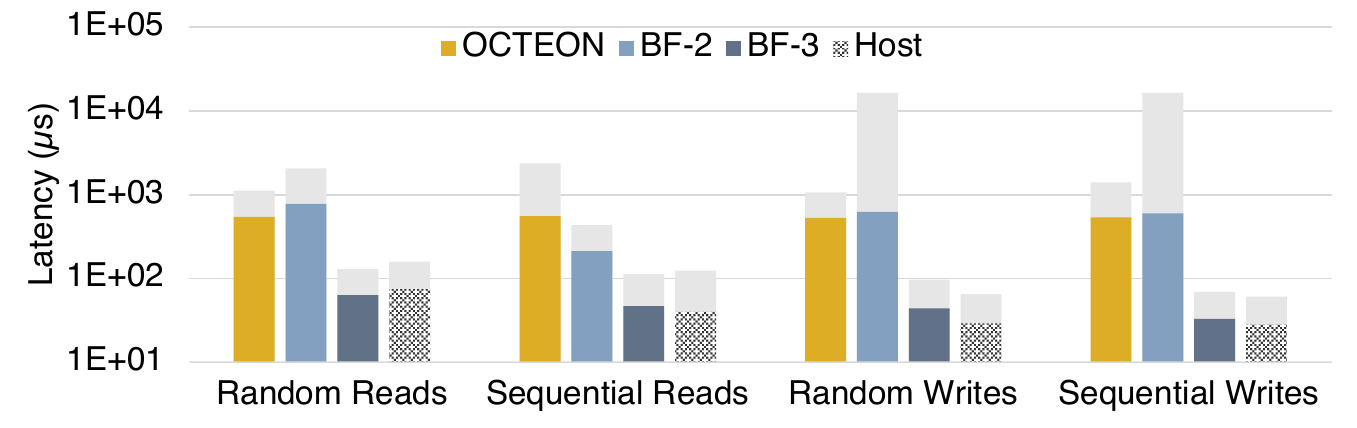}
        \caption{Latency of 8\,KB accesses.}
        \label{fig:stor-lat-8kb}
    \end{subfigure}%
    \hfill
    \begin{subfigure}{\columnwidth}
        \centering
        \includegraphics[width=\textwidth]{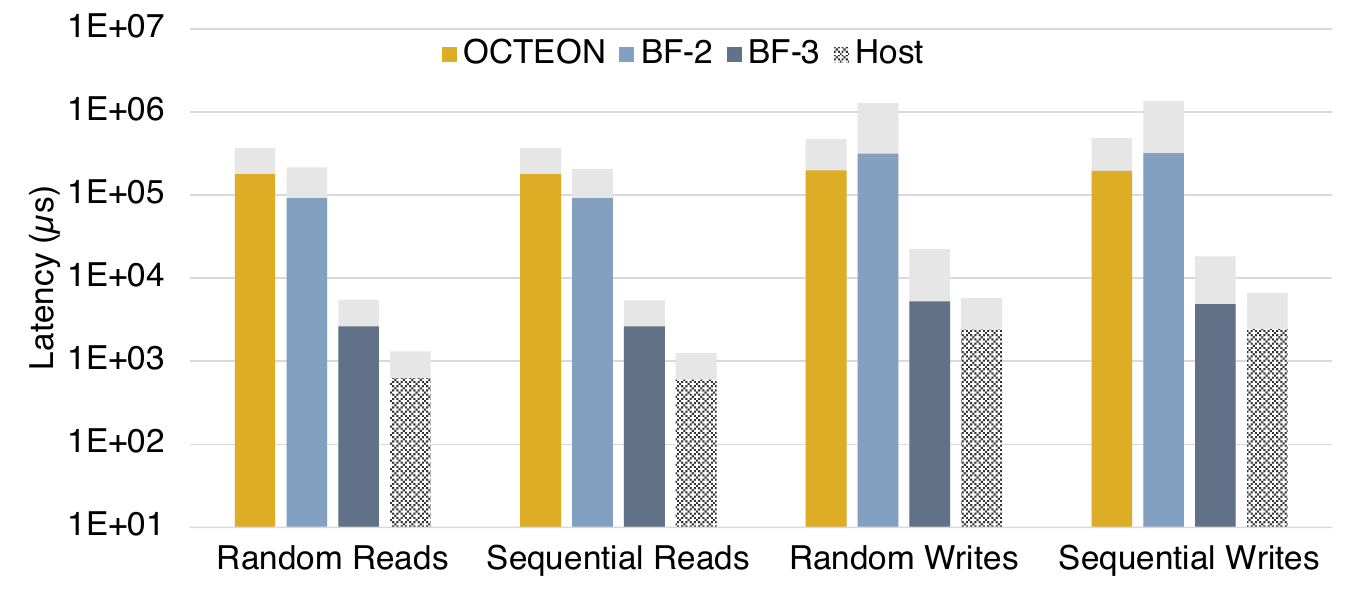}
        \caption{Latency of 4\,MB accesses.}
        \label{fig:stor-lat-4mb}
    \end{subfigure}
    \caption{Benchmarking storage latency. Foreground bars of different colors represent average latency, and background light grey bars represent p99 tail latency.}
    \label{fig:stor_lat}
\end{figure}

We first tune the above parameters on each DPU and the host to achieve its highest storage I/O throughput.
The results are shown in Figure~\ref{fig:stor_thr}.
Across all settings, we observe three levels of performance: the slowest eMMC flash devices on \octeon and \nvbf (10s--100s MB/s), the faster NVMe SSD on \nvbff (100s--1000s MB/s), and the best-performing NVMe SSD on the host (1000s MB/s).
While the NVMe SSD on \nvbff is much faster than the storage on other DPUs, the gap to the host storage device remains large---2.8$\times$--10.5$\times$ slower.
The host storage performance is two orders of magnitude higher than that on the slower DPUs.
The impact of storage testing parameters varies across platforms.
When performing random reads, increasing the access size from 8\,KB to 4\,MB reduces the degree of randomness (since bytes within an access are sequentially read from the device), which is more beneficial for \nvbf and \nvbff than for \octeon and the host (350\% and 440\% vs. 50\% and 150\% throughput increase, respectively).
When changing the access pattern completely from random reads to sequential reads, the throughput of \nvbf increases significantly for small accesses: a 250\% increase for 8\,KB reads.
The benefit is not reflected in other platforms: only a 17\% difference is observed on the host, which shows the efficiency of SSDs in random accesses.
Writes are generally slower than reads (Figures~\ref{fig:stor_write_rnd} and~\ref{fig:stor_write_seq} vs. Figures~\ref{fig:stor_read_rnd} and~\ref{fig:stor_read_seq}).
Similarly to reads, \octeon and \nvbf are slower than \nvbff and the host in writes, and different access sizes and patterns impose the highest impact on \nvbf.
The gap between \nvbff and the host is larger than in reads.

Figure~\ref{fig:stor_lat} shows the storage I/O latency.
In this experiment, we set the number of outstanding requests and the number of threads both as one to achieve the lowest latency on each platform.
Figures~\ref{fig:stor-lat-8kb} and~\ref{fig:stor-lat-4mb} report the average and tail latencies of the 8\,KB and 4\,MB accesses, respectively.
We make more promising observations than in the throughput results: the latency of DPUs (particularly \nvbff) can be on par with and even lower than that of the host.
For small random and sequential reads (Figure~\ref{fig:stor-lat-8kb}), the tail latency of \nvbff is $\sim$20\% lower compared to the host.
Its average latency is also lower in serving random reads.
This is especially advantageous for serving remote storage requests considering the shorter distance to the network interface on the DPU from the DPU storage device.
However, when accessing larger blocks (4\,MB), where performance becomes bandwidth-bound, Figure~\ref{fig:stor-lat-4mb} shows that even on \nvbff, the time is 3$\times$--5$\times$ higher than on the host.

Although directly attached storage devices on DPUs are auxiliary and do not target on-path application offloading, we identified scenarios where storage offloading can be helpful.
Our findings from the storage benchmarks are summarized below.

\vspace{3mm}
\begin{mdframed}[style=FindingFrame,nobreak=false,align=center,userdefinedwidth=27em]

\begin{itemize}
    \item {\em DPUs generally provide considerably lower storage performance than the host for throughput-bound and large I/Os.}
    \item {\em The latest DPU achieves low latency for fine-grained accesses.}
\end{itemize}

\end{mdframed}

\subsection{Networking}

The final microbenchmark focuses on the performance of moving data across the network.
In this experiment, we set up a physical network in which a remote server (with the same configuration as the host machine) connects to one of our DPUs (\nvbf) via a 100\,Gbps cable.
The testing parameters are specified as follows. 
We measure both the network transfer latency and throughput. 

\begin{table} [h]
\small
\setlength{\tabcolsep}{10pt}
    \centering
    \begin{tabular}{c c c}
        \toprule
        {\bf Message size} & {\bf Queue Depth}  & {\bf \#Threads} \\
        \midrule
        32\,B -- 32\,KB &
        1--128 &
        1--Max\\
        \bottomrule
    \end{tabular}
    \label{tab:network-bench-setup}
\end{table}

\begin{figure}[t]
    \centering
    \begin{subfigure}{.49\columnwidth}
        \centering
        \includegraphics[width=\textwidth]{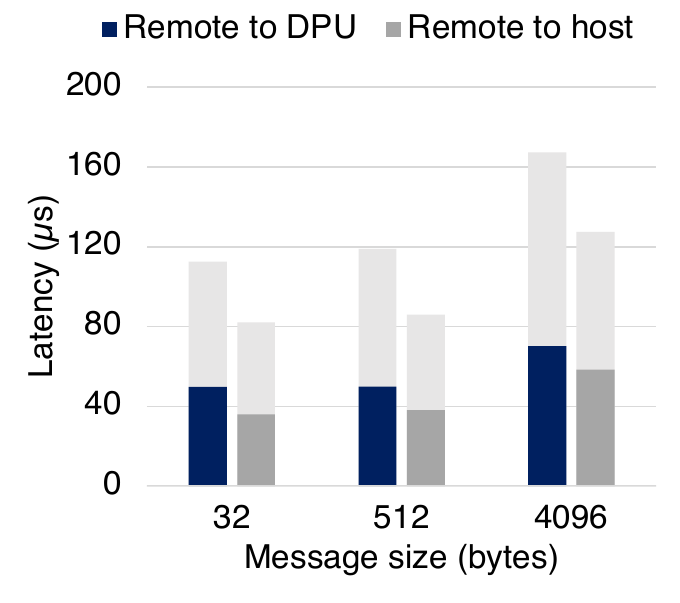}
        \caption{Latency (average and p99).}
        \label{fig:net-tcp-lat}
    \end{subfigure}%
    \hfill
    \begin{subfigure}{.49\columnwidth}
        \centering
        \includegraphics[width=\textwidth]{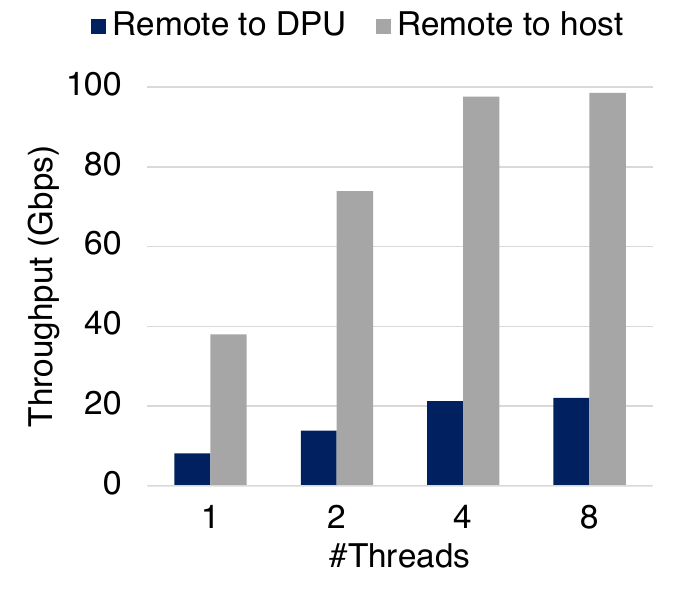}
        \caption{Throughput.}
        \label{fig:net-tcp-thr}
    \end{subfigure}
    \caption{Benchmarking network performance.}
    \label{fig:stor_lat}
\end{figure}

We first investigate the network latency.
To do so, we spawn one thread on the remote server, which issues closed-loop ping-pong messages to accurately measure the network round-trip latency.
We compare the latency between the remote server and the DPU to that between the remote server and the local host.
This comparison seeks to assess the efficiency of the DPU for communication offloading.
Figure~\ref{fig:net-tcp-lat} shows the average and tail latencies of different message sizes.
We observe that generally the latency between the remote server (``remote'' for short) and the DPU is higher than that between remote and the host on all message sizes.
On average, the former is 30\% higher than the latter.
We ascribe this latency overhead on the DPU to its wimpy CPU, where the Linux TCP/IP stack runs---software simply becomes slower.

We next measure network throughput, where  the above finding is also reflected. 
For this measurement, we let the remote server and the DPU/host create multiple threads, each processing a connection for exchanging large messages (32\,KB).
Each connection maintains a queue depth of 128 to ensure that the connection throughput is saturated.
Figure~\ref{fig:net-tcp-thr} reports the network throughput between remote and the DPU and between remote and the host.
The trend is clear---the performance gap to the host is larger than in latency tests.
With one thread, the DPU can achieve 8\,Gbps throughput, while the single-thread throughput of the host is 4.8$\times$ higher at 38 \,Gbps.
Both the DPU and the host saturate reach the peak throughput with four threads, where the DPU's throughput is 22\,Gbps and the host's throughput is 98\,Gbps.
In fact, the single-thread throughput of the host is 1.7$\times$ higher than the eight-core throughput of the DPU. 
This result shows that offloading high-throughput network communication to the DPU can cause great performance degradation {\em using the TCP/IP stack in the onboard Linux}.

\begin{figure}[t]
    \centering
    \begin{subfigure}{.49\columnwidth}
        \centering
        \includegraphics[width=\textwidth]{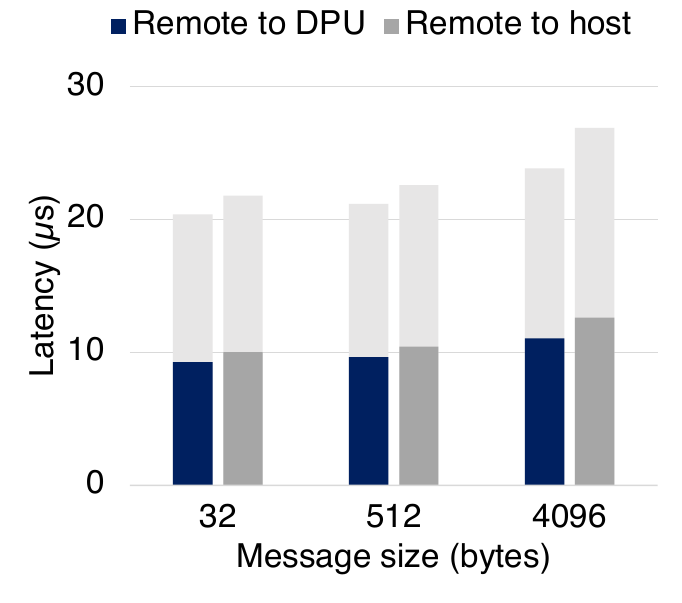}
        \caption{Latency (average and p99).}
        \label{fig:net-rdma-lat}
    \end{subfigure}%
    \hfill
    \begin{subfigure}{.49\columnwidth}
        \centering
        \includegraphics[width=\textwidth]{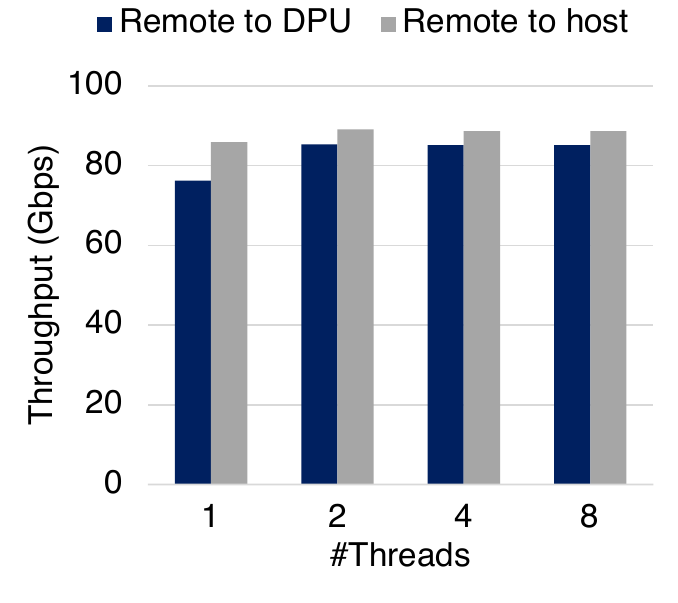}
        \caption{Throughput.}
        \label{fig:net-rdma-thr}
    \end{subfigure}
    \caption{Network performance with RDMA.}
    \label{fig:stor_lat}
\end{figure}

To verify that the above inefficiencies originate from the software stack running on the weaker cores, we implemented an additional plugin task in \name for  network benchmarking with Remote Direct Memory Access (RDMA) over InfiniBand, supported by the \nvbf DPU.
RDMA bypasses the onboard Linux to directly access the NIC for data transfers.
Specifically, the task uses NVIDIA-provided {\tt ib\_read\_lat} and {\tt ib\_read\_bw} tools to perform RDMA reads from the remote server to the DPU/host and measure network performance.
We use the same parameters provided in the previous latency and throughput measurements.
Figures~\ref{fig:net-rdma-lat} and~\ref{fig:net-rdma-thr} compare the kernel-bypass latency and throughput between remote and the DPU to that between remote and the host. 
We observe that when the networking stack in the onboard OS is bypassed, {\em network communication to the DPU has lower latency than to the host}.
As Figures~\ref{fig:net-rdma-lat} shows, when accessing 4\,KB data, the latency of the DPU is 12.6\% lower than that of the host.
The lower latency can be explained by the shorter distance from the NIC to the DPU memory than to the host memory.  
Regarding throughput, although the single-connection performance of the DPU is still lower than the host, the gap is marginal (11.3\%).
The peak throughput is achieved with two threads/connections for both the DPU and the host, where the throughput gap is further closed.

Below we summarize insights into DPU networking.

\vspace{3mm}
\begin{mdframed}[style=FindingFrame,nobreak=false,align=center,userdefinedwidth=27em]

\begin{itemize}
    \item {\em Offloading network communication to DPUs with the onboard TCP stack reduces performance, especially throughput.}
    \item {\em Kernel-bypass networking can eliminate the impact of weak cores and {\bf\em achieve even lower latency than the host}.}
\end{itemize}

\end{mdframed}

\section{Benefits of DPU Offloading}
\label{sec:module}
\rev{We now investigate the benefits of near-data processing and augmented processing capabilities enabled by DPUs, and present the results of offloading the two cloud database modules included in \name on different DPUs.}

\subsection{\rev{Predicate Pushdown}}
\label{sec:predpush}

\begin{figure}
    \centering
    \includegraphics[width=\columnwidth]{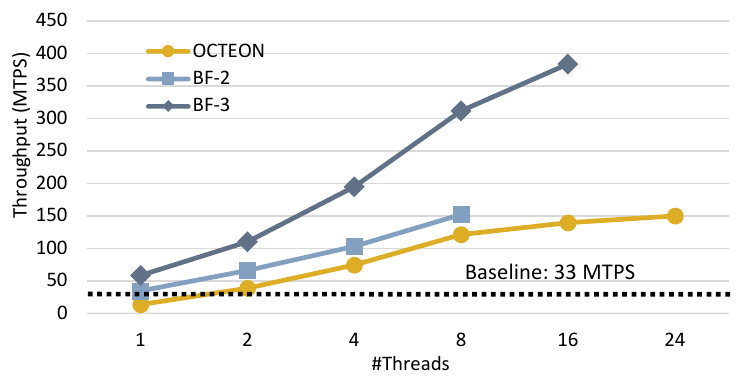}
    \caption{Predicate pushdown}
    \label{fig:pushdown}
\end{figure}

\rev{We fix the scale of the database as 10\,GB (TPC-H scale factor 10) and the predicate selectivity as 1\%, and vary the number of cores utilized on the DPU for the scan from 1 to all available cores.}

\rev{Figure~\ref{fig:pushdown} shows the performance of the basic scan and DPU-enabled predicate pushdown. The baseline where the entire table is fetched from the storage server to the compute server incurs expensive network and storage I/O and thus has low scan throughput---33 million tuples per second (MTPS). Pushing down the predicate to the DPU on the storage server can eliminate most of the data movement: the weaker DPUs, i.e., BF-2 and OCTEON, outperform the baseline when two cores are utilized for the scan, and when all their DPU cores are utilized, their throughput reaches 150 MTPS, 4.5$\times$ faster than the baseline. BF-3 is significantly faster than other solutions. Its speedup over the baseline is 1.8$\times$ with a single core and 12$\times$ with all its 16 cores. This result confirms the potential of DPUs for near-storage data processing.}

\subsection{\rev{Index Offloading}}
\label{sec:indoff}
\begin{figure}
    \centering
    \includegraphics[width=\columnwidth]{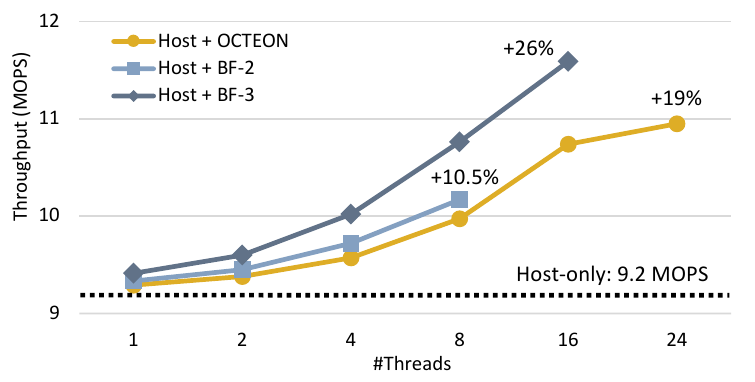}
    \caption{Offloading index operations to DPUs.}
    \label{fig:indoff}
\end{figure}

\rev{To evaluate the performance increase when using a DPU as a coprocessor of the host, we set up an index of 50\,GB (50 millions of 1\,KB records) and split it between the host and the DPU with a ratio of 10:1. We then execute uniform reads on the host and the DPU separately and measure the overall index throughput. Recent work has reported the performance increase by sharing index workloads between the host and the BF-1 and BF-2 DPUs~\cite{ThostrupFZB22}, our experiments further examine this benefit on a more recent DPU (BF-3) and a DPU from a different brand (OCTEON). The results are shown in Figure~\ref{fig:indoff}. Without any offloading, the host can achieve 9.2 million operations per second (MOPS) with 96 threads. By offloading part of the index to the DPU, more requests can be serviced per time unit. Although each individual DPU core is weaker than the host CPU, fully utilizing the DPU leads to noticeable performance increase: 19\%, 10.5\%, and 26\% higher throughput when offloading to OCTEON, BF-2, and BF-3, respectively.}

\section{End-to-end DBMS Performance}
\label{sec:e2e}

\rev{We finally benchmark the DPUs with the built-in DBMS task of \name. Note that explicitly not our goal is advocating full system deployment on DPUs. Rather, our results recognize the overhead of doing so and thus motivate co-designs between data systems and DPUs, which is aligned with recent proposals incorporating partial offloading~\cite{dds, smartshuffle}.}
\rev{On each DPU platform, we allocate all available cores to DuckDB and measure its running time for each TPC-H query (scale factor 10) under cold and hot executions.}

\begin{figure}[t]
    \centering
    \begin{subfigure}{\columnwidth}
        \centering
        \includegraphics[width=\textwidth]{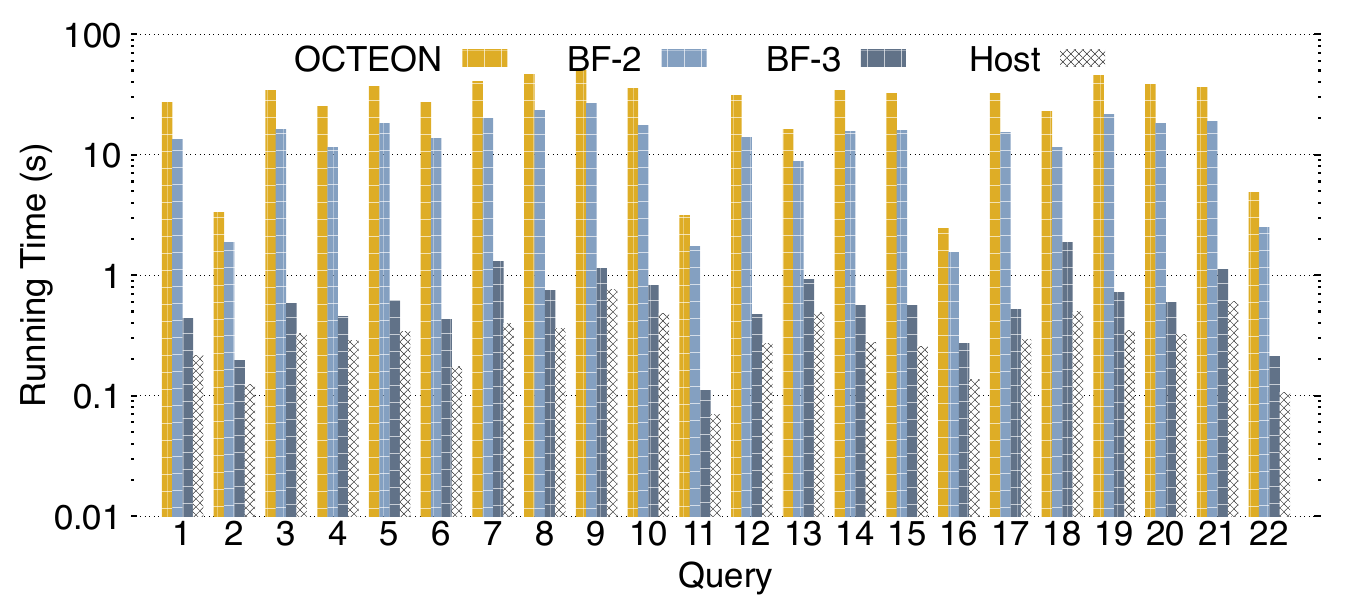}
        \caption{Cold execution.}
        \label{fig:dbms-cold}
    \end{subfigure}%
    \hfill
    \begin{subfigure}{\columnwidth}
        \centering
        \includegraphics[width=\textwidth]{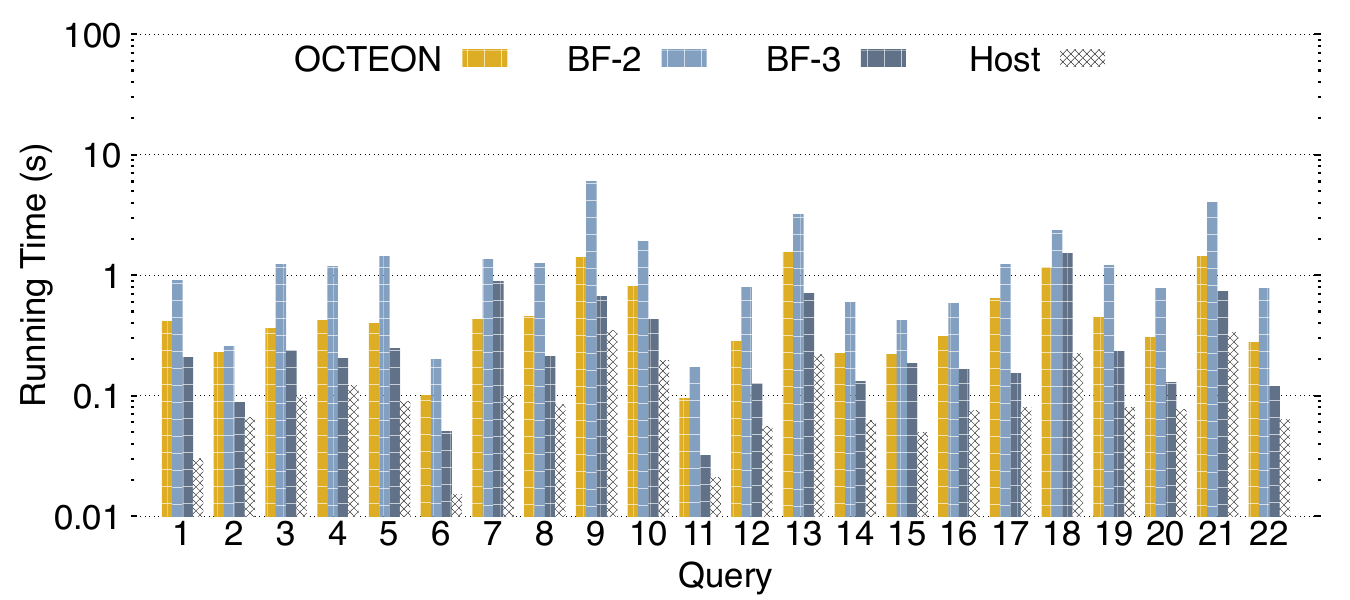}
        \caption{Hot execution.}
        \label{fig:dbms-hot}
    \end{subfigure}
    \caption{Running times of DuckDB on different platforms.}
    \label{fig:dbms}
\end{figure}

Figure~\ref{fig:dbms-cold} shows the results of cold executions.
The host outperforms all DPUs, as expected.
Its average query execution performance is 87$\times$, 43$\times$, and 2.1$\times$ higher than \octeon, \nvbf, \nvbff, respectively.
The primary bottleneck in this execution mode is disk I/O, particularly sequential reads as the tables are scanned and loaded into the main memory.
Recall from Section~\ref{sec:stor} that the eMMC flashes on \octeon and \nvbf are much slower than the NVMe SSDs on \nvbff and the host, which is reflected in the end-to-end query execution.
Between the DPUs, \nvbff is 21$\times$ faster than its previous generation, and as Figure~\ref{fig:stor_read_seq} shows, \nvbf achieves higher sequential read performance, and thus the average query processing time on \nvbf is 2$\times$ shorter than on \octeon.

We make different observations with hot executions, which are illustrated in Figure~\ref{fig:dbms-hot}.
Since disk I/O is avoided, 
the performance of CPU and memory now dominates query execution.
The gap between the host and the best-performing DPU, i.e., \nvbff, (3$\times$) increases compared to cold executions.
This can be attributed to CPU and memory differences.
In particular, the DPU has much less cores (16 on \nvbff vs. 96 on the host).
It also explains the change between \octeon (24 cores) and \nvbf (8 cores).
The former, which was slower in cold executions, is now 2.7$\times$ faster.

\rev{In summary, when running a full DBMS, storage performance, CPU performance, core count, and memory efficiency together create a significant gap between the host and the DPUs.}

\section{Related Work}
\label{sec:related}

{\em DPU Benchmarking.}
DPUBench~\cite{dpubench} proposes a 
benchmark suite that targets DPUs.
The benchmark suite includes operators for storage, 
networking, 
and security. 
It also includes end-to-end applications that use the operators for communications, file compression and integrity checking, and security.
However, DPUBench only evaluates \nvbf.
Even though DPUBench shares some operator benchmarks with \name, it 
is insufficient to evaluate data processing systems, which is the major contribution of \name.

DPU-Bench~\cite{dpu-bench} evaluates DPU performance under HPC scenarios.
Specifically, the benchmark suite measures 
RDMA-based MPI communication to \nvbf.
The goal is to determine the number of DPU processes to maximize offloading efficiency.
The follow-on work~\cite{DBLP:conf/hoti/MichalowiczSSPP23} compares \nvbf and \nvbff. 
\name, in contrary, focuses on DPU's capability to offload data processing tasks.

Wei et al.~\cite{bf2_rdma} quantitatively analyzed the performance of various RDMA paths on \nvbf. 
Based on the benchmark studies, the work proposes an optimization guide, and applies the guide to prior DPU-accelerated file system and key-value store.
Chen et al.~\cite{DBLP:journals/corr/abs-2402-03041} performed a more comprehensive performance studies, including both compute and communication, of \nvbff, with a focus on the RISC-V data path accelerators (DPA).
Unlike \name, both works study only a single DPU platform;
their benchmarks are also inadequate to directly evaluate data processing applications.

Other prior works benchmark specific aspects of DPUs.
Li et al.~\cite{DBLP:journals/micro/LiKGL24} evaluate the performance of lossy (SZ3) and lossless (DEFLATE, lz4, zlib) compression in the SoC and ASIC ("C-engine") implementations in \nvbf and \nvbff.
Liu et al.~\cite{bf2_perf_char} use the {\tt stress-ng} tests to evaluate \nvbf. 
Zhou et al.~\cite{zhou2022} measure the performance of offloading microservices to an Intel IPU.
Compared to these works, \name provides a more comprehensive benchmark suite for heterogeneous DPU platforms.
\name also 
targets offloading for data processing systems, which presents a gap in the literature.

\vspace{2mm}
\noindent{\em DPU Offloading.}
DPU has been a popular hardware target for application and system offloading.
LineFS~\cite{linefs} offloads operations in a distributed file system to BlueField-2.
Xenic~\cite{xenic} partially offloads data store and concurrency control to accelerate a distributed transactional system.
Several prior systems~\cite{fairnic,flextoe,acceltcp} demonstrate the performance benefit of offloading network functions to DPUs.
A line of work~\cite{ipipe,floem} designs general programming frameworks to offload applications to DPUs.
\name, in comparison, targets offloading data processing system components to DPUs. 

\section{Conclusion}
\label{sec:conclusion}

We first presented \name, a benchmark framework for measuring data processing performance on DPUs.
It provides an extensible task abstraction, on top of which a variety of performance tests can be incorporated and automated by the framework.
With \name, we implemented a set of microbenchmarks, which measure the CPU, memory, networking, and storage performance of DPUs, \rev{the offloading of two cloud database modules, as well as a full DBMS}.
This benchmark suite has been used to measure the performance of recent DPUs from NVIDIA and Marvell.
Based on the results, we provided useful insights into the performance characteristics of DPUs for offloading data processing tasks, from primitive operations to an end-to-end system.


\balance
\bibliographystyle{abbrv}
\bibliography{ref}

\end{document}